\documentclass[preprintnumbers,floatfix,aps]{revtex4-1}
\usepackage{amsmath}
\usepackage{amsfonts}
\usepackage{amssymb}
\usepackage{graphicx}
\usepackage{epsfig}
\usepackage[hang,nooneline]{subfigure}
\usepackage{color}

\newcounter{fig}

\usepackage{graphicx}
\usepackage{epsfig}
\usepackage[hang,nooneline]{subfigure}
\usepackage{color}
\newcommand{\pstar}{\mbox{$\psi^{\ast}$}}
\newcommand{\bq}{\begin{equation}}
\newcommand{\ee}{\end{equation}}
\newcommand{\bea}{\begin{eqnarray}}
\newcommand{\eea}{\end{eqnarray}}

\newcommand{\ba}{\begin{eqnarray}}
\newcommand{\eq}{\end{equation}}
\newcommand{\ea}{\end{eqnarray}}

\newcommand{\sech}{{\rm sech}}

\begin{document}
\preprint{LA-UR 16-23477, \today}

\title{Variational Approach to studying solitary waves in the nonlinear Schr\"odinger equation with Complex Potentials}
\author{Franz G. Mertens}
\email{Franz.Mertens@uni-bayreuth.de}
\affiliation{Physikalisches Institut, Universit{\"a}t Bayreuth, D-95440 Bayreuth, Germany} 

\author{Fred Cooper}\
\email{cooper@santafe.edu}
\affiliation{Santa Fe Institute, Santa Fe, NM 87501, USA}
\affiliation{Center for Nonlinear Studies and Theoretical Division, Los Alamos National Laboratory, Los Alamos, New Mexico 87545, USA}

\author{Edward Ar\'evalo}
\email{earevalo@fis.puc.cl} 
\affiliation{Pontifical Catholic University of Chile,
Departamento de F\'{\i}sica, Santiago, Regi\'on Metropolitana, Chile}

%\author{Niurka R. Quintero} 
%\email{niurka@us.es} 
%\affiliation{IMUS and Departamento de Fisica Aplicada I, E.P.S. Universidad de Sevilla, 41011 Sevilla, Spain}

\author{Avinash Khare}
\email{khare@physics.unipune.ac.in} 
\affiliation{Physics Department, Savitribai Phule Pune University, Pune 411007, India}

\author{Avadh Saxena} 
\email{avadh@lanl.gov} 
\affiliation{Center for Nonlinear Studies and Theoretical Division, Los Alamos National Laboratory, Los Alamos, New Mexico 87545, USA}
\author{A.R. Bishop} 
\email{arb@lanl.gov} 
\affiliation{ Los Alamos National Laboratory, Los Alamos, New Mexico 87545, USA}

\date{\today}

\begin{abstract}
We discuss the behavior of solitary wave solutions of the  nonlinear Schr{\"o}dinger equation (NLSE)  as they interact with complex potentials, using a four parameter variational approximation based on a dissipation functional formulation of the dynamics.  We concentrate on spatially periodic potentials with the periods of the real and imaginary part being either the same or different. Our results  for the time evolution of the collective coordinates of our variational ansatz are in good agreement with direct numerical simulation of the NLSE.  We compare our method with a collective coordinate approach of Kominis and give examples where the two methods give qualitatively different answers.  In our variational approach, we are able to give analytic 
results for the small oscillation frequency of the solitary wave oscillating parameters which agree with the numerical solution of the collective coordinate equations. 
We also verify that  instabilities set in when the slope of $dp(t)/dv(t)$ becomes negative when plotted parametrically as a function of time, where $p(t)$ is the momentum of the solitary wave and $v(t)$ the velocity. 
\end{abstract}

\maketitle
\section{Introduction}
The behavior of solitary waves in the presence of complex potentials and in particular solitary wave solutions of the  NLSE in the presence of complex potentials has been the subject of much recent investigation \cite{musslimani}  \cite{PT_periodic}  \cite{kominis} \cite{kev}.  Complex  potentials  in Quantum Mechanics  with $\mathcal{PT}$ symmetry  \cite{Bender_review,special-issues,review} possess special properties such as having real spectra.   By further imposing other relations such as solvability as a result of supersymmetry one can restrict the behavior  of  solitary waves which occur when we add these potentials to the NLSE \cite{kominis}.  In a recent paper, some of us studied the  behavior of the exact solitary wave solutions of the NLSE in the presence of a complex $\mathcal{ PT}$ symmetric trapping potential \cite{kev}.
 Motivated
to a considerable degree by the study of the specially balanced
$\mathcal{PT}$-symmetric dynamical
models~\cite{Bender_review,special-issues,review},  there has been, in the past 15 years, a large number
of studies of open systems having both gain and loss. 

The original proposal of Bender and collaborators to 
study  such systems was made as an alternative to the
postulate of hermiticity in quantum mechanics. Yet, in the
next decade, proposals aimed at the experimental realization
of such $\mathcal{PT}$-symmetric systems found a natural setting
in the realm of optics~\cite{Muga,PT_periodic}. Within the latter, the above
theoretical proposal (due to the formal similarity of the Maxwell
equations in the paraxial approximation and the nonlinear  Schr{\"o}dinger
equation) quickly led to a series of experiments~\cite{experiment}.   As noted in  \cite{musslimani} $\mathcal{PT}$ symmetric behavior should be observable
in standard quantum well semiconductor lasers or semiconductor optical amplifiers \cite{exp2}.
The possibility of experimentally observing the effects of $\mathcal{PT}$ symmetry has motivated experiments in numerous other areas,
which  include the examination of
$\mathcal{PT}$-symmetric electronic
circuits~\cite{tsampikos_recent,tsampikos_review}, mechanical
systems~\cite{pt_mech} and whispering-gallery microcavities~\cite{pt_whisper}.  In all these systems solitary waves play an important role
in the dynamics and the behavior of the solitary waves in these complex potentials can now be explored experimentally.
Thus having a  simple way of examining the dynamics of these solitary waves and their stability properties is quite important for future experiments.  

Our paper is organized as follows.  In section II we introduce a generalized variational method for obtaining the NLSE in the presence of complex potentials. 
This requires the introduction of a dissipation functional.  In that section we also show how to introduce macroscopic  collective variables, which  depend only on  time, based on the real density and current  familiar from the Schr{\"o}dinger equation. The dynamics of these macroscopic variables depend on integrals over the 
real and imaginary parts of the external potential. These results just  depend on assuming  the solitary wave
wave-function $\psi$  is  a function of $x-q(t)$ and is a solution of the NLSE   in the presence of an external complex potential.   In section III, we make use of the variational formulation of the dynamics to introduce a reduced parameter space approximation to the dynamics, where the dynamics is obtained from
the variational principle which includes a dissipation function. In this approach we parametrize the solitary wave function by collective variables representing the amplitude, width, position, and phase of the solitary wave. We have successfully used this four collective
coordinate approach (4 CC)   earlier \cite{mert1} \cite{mert2} \cite{coop1}  in studying the effect of external forces in the NLSE.  There we found the stability criterion
\bq \label{stab1}
\frac{dp}{d v} ~ >0,
\eq
where $p(t), v(t) = \dot q(t)$ are a parametric representation of the curve $p(v)$.    
Here  $p(t)=P(t)/M(t)$, is the scaled momentum of the soliton, and $v(t) = \dot q(t)$ is the velocity of  of the solitary wave. We will define  these variables more precisely
below.  What we found in our previous studies  \cite{mert1} \cite{mert2} \cite{coop1} , was that whenever Eq. 
\eqref{stab1} was violated anywhere on the curve $p(v)$ the soliton became unstable. i.e. $dp/dv < 0$ is a sufficient condition for instability. We will show in what follows
that this instability either leads to the solitary wave then oscillating at another frequency, or blowing up or collapsing.  The usefulness of this criterion for studying soliton stability in generalized NLSEs was investigated in detail in \cite{niurka}. 
%For the case of the ``pinned" solitary wave trapped in a PT symmetric Scarf II potential that we studied earlier numericaly, we will need to introduce
%a "chirp" term to the equations so that both the "Derrick" type of instability applied to the ``width" of the solitary wave as well as instabilities in the position of the solitary wave can be studied. 
In section IV we discuss the simplification of the collective coordinate dynamics that occurs when the complex external potential is $\mathcal{PT}$ symmetric. In Section V we briefly describe our method of numerically solving the NLSE.  In section VI we consider several examples of complex potentials previously considered by Kominis \cite{kominis} , in order to compare our approach to his and also to direct numerical simulations.  In section VI we summarize our main conclusions. 

\section{Dissipation functional Formulation of the NLSE with a Complex potential}
We are interested in devising a variational principle for obtaining the equation for the wave function and its complex conjugate for the NLSE in a complex potential.  The complexity of the potential makes the problem non-conservative and there are several approaches to dealing with this problem.  Here we will use an extension of the Dissipation Functional method that we used previously \cite{coop1} when the complex part of the potential was a constant. 

The equations we are interested in studying are:
\bq \label{NLSE}
 i \psi_{t} + \partial_x^{2}\psi +  g (\pstar\psi)^{\kappa} \psi - (V+ i W ) \psi= 0 
\ee
as well as its complex conjugate equation:
\bq \label{CNLSE}
 - i \psi_t^\star + \partial_x^{2}\pstar +  g(\pstar\psi)^{\kappa} \pstar- (V- i W ) \pstar= 0 . 
\ee

Let us define the usual conservative part of the action as
\bq
\Gamma = \int dt L_c ,  \label{eq:gamma-definition}
\ee
where the conservative part of $L_c$ depends only on the real part of the potential and is given by

\bq
L = \int {\cal L} dx \,  =  {\frac{i}{2}} \int d x (\pstar \psi_{t} -\psi_t^\star\psi) - H_c . \label{eq:action}
\ee
For the NLSE  with arbitrary nonlinearity parameter $\kappa$ in $d$ spatial dimensions we have

\bq
H_c  = \int d x [\partial_x\pstar \partial_x\psi - g
{\frac{(\pstar\psi)^{\kappa+1}}  {\kappa+1}} + \pstar V(x) \psi] . 
\ee

We    will introduce the Dissipation Functional  $F$ via
\bq
F = \int {\cal F}  d x dt,
\eq
where
\bq
{\cal F} =  i W(x) \left( \psi_t \psi^\star -\psi_t^\star \psi \right).
\eq
The  equations for the wave function of the NLSE in the presence of  a  complex potential follow from the generalized Euler-Lagrange Equations: 
\bq \label{nlse}
\frac{\delta \Gamma}{\delta \pstar} = - \frac{\delta F}{\delta\psi_t^\star }
\eq
and its complex conjugate equation.  Equation \eqref{nlse}  leads to 
\bq
\partial_t \frac{\partial {\cal L} }{\partial\psi_t^\star} +\partial_x \frac{\partial {\cal L} }{\partial\psi_x^\star} -\frac{ \partial {\cal L}} {\partial \pstar} = \frac{ \partial {\cal F}} {\partial\psi_t^\star},
\eq
which yields  Eq. \eqref{NLSE}.
%\bq i \psi_{t} + \partial_x^{2}\psi +  g (\pstar\psi)^{\kappa} \psi - (V+ i W ) \psi= 0 
%\ee
The complex conjugate of  Eq. \eqref{nlse}  leads to the complex conjugate of the NLSE equation, namely Eq. \eqref{CNLSE}.
%\bq - i\psi_t^\star + \partial_x^{2}\pstar +  g(\pstar\psi)^{\kappa} \pstar- (V- i W ) \pstar= 0 
%\ee

If we multiply Eq. \eqref{NLSE}  by $\pstar$  and add the complex conjugate, then W drops out from the resulting equation and
we can obtain a Virial Theorem for the spatial average of the  potential. Explicitly we find: 
\bq
{\frac{i}{2}} \int d x (\pstar \psi_{t} -\psi_t^\star\psi) - \int d x  \left [\partial_x\pstar \partial_x\psi - g
{\frac{(\pstar\psi)^{\kappa+1}}  {\kappa+1}}\right] = \int d x  \pstar V(x) \psi].
\eq
Another approach for handling complex potentials has been recently put forth by Rossi et al  \cite{Rossi}. However, for the problem at hand the dissipation function is sufficient as it leads to results consistent with the equations for the single particle variables of the soliton, such as mass, position and momentum  derived directly from the NLSE, as discussed in the next subsection.
In what follows we will be interested in the particular case $\kappa=1, g=2$. ($g$ of course can be scaled out of the NLSE by a rescaling of the fields). 
%{\it Note that this agrees with the usual SE convention for the sign of $V$ and differs from Kominis}.

\subsection{General Properties of the NLSE in  complex potentials}
A general approach for studying soliton dynamics has been discussed for real potentials in the work of
Quintero, Mertens and Bishop \cite{niurka}  and also by Kominis \cite{kominis}  for complex potentials. Here we follow the approach of \cite{niurka}.
We are interested in  solitary wave solutions that approach zero exponentially  at $\pm \infty$.  For these solutions we define the mass density $\rho(x,t) = \pstar \psi$, and 
the mass or norm  $M(t)$  as
\bq
M(t) =\int dx \,  \rho(x,t)  =  \int dx \,   \pstar(x,t) \psi(x,t).
\eq
We also define the current as: 
\bq
j(x,t) = i (\psi\psi_x^\star - \pstar \psi_x).
\eq
From the NLS equations we have
\bq \label{cont}
\frac{ \partial \rho(x,t)} {\partial t} +  \frac{\partial j(x,t) }{\partial x}  =  2 W(x) \rho(x,t).
\eq
Integrating over space, and assuming that $\rho(x,t) = \rho(x-q(t),t) $,  where $q(t)$ is the position of the solitary wave,
we find 
\ba\label{mdot}
\frac{dM(t)}{dt} &&=  2 \int dx \,  \, W(x) \, \rho (x-q(t),t)  \\
&&=2 \int dy \,  W[y+q(t) ] \,  \rho(y,t) .
\ea
Here the explicit time dependence in $ \rho(x-q(t),t) $ takes into account that the shape of the soliton may depend on time. 
We observe that  $M$ is conserved when $W(x) =0$. 
From this we have that when $W(x) = - \alpha$, with $\alpha$ a positive constant,  the mass dissipates to zero, since
\bq \label{dissM}
\frac{dM(t)}{dt} = - 2 \alpha M(t)  \rightarrow M(t) = M(0) e^{- 2\alpha t}.
\eq

If $\rho(x,t)$ is symmetric about its midpoint, then $q(t)$ can be defined through:
\bq
M(t) q(t) = \int dx \,  ~  x ~ \pstar(x-q(t),t ) \psi(x-q(t),t ).
\eq

Multiplying the continuity equation Eq. \eqref{cont}  by $x$ and integrating over all space we find:
\bq
 \frac{d}{dt} \int dx \,  ~  x \,  \rho(x,t) = 2 P(t) + \int dx \,  \,  \left( 2 x \,  W(x) \, \rho(x,t) \right),
\eq
where \bq
P(t) =  \frac{1}{2} \int dx \,   \, j(x)= \int dx \,  \, \left[ \frac{i}{2} \left( \psi\psi_x^\star - \pstar \psi_x \right) \right].
\eq
Again assuming $\rho(x,t) = \rho(x-q(t),t) $, we can write this equation, using $y=x-q(t)$,   as:
\bq
\frac{d}{dt} \left[ M(t) \,  q(t)  \right] =2 \,  P(t) + \int dy\,  2  \, y  \, W[y+q(t)] \, \rho(y,t) +2  \, q(t) \, \int dx \,   \, W[x]  \, \rho(x-q(t),t) .
\eq
We recognize the last term as $q(t) dM(t)/dt$, so that we finally have:
\bq
M(t) \, \frac{dq(t)}{dt} = 2 \,  P(t) + \int dy  \, 2 \,  y  \, W[y+q(t)]  \, \rho(y,t) .
\eq
Letting $p(t) = P(t)/M(t)$, we  obtain
\bq \label{litp}
 \dot q(t) = 2 p(t) + \frac{1}{M(t)} \,  \int dy \, 2\, y \, W[y+q(t)]  \, \rho(y,t).
\eq
Taking the time derivative of the momentum $P (t)$, using the equations of motion for $\psi$  and 
$\pstar $  and integrating by parts, we find

\bq \label{dpdt}
\frac{dP}{dt}  = - \int dx \,   \, \rho (x,t) \frac{ \partial V}{\partial x}  +  \int dx \,  ~ j(x,t ) \,  W(x). 
\eq

Assuming as in quantum mechanics that
\bq
\frac {1}{i} \frac{\partial }{\partial x} \psi (x, t) = p(t) \, \psi(x,t ) ,
\eq
or equivalently
\bq
 \frac{1}{2} j(x, t)  = p(t)  \,  \rho (x, t) ,
\eq
 which is the local version of the integral relationship $ P(t) = M(t) p(t)$, then the last term in Eq. \eqref{dpdt} is $p \, dM/dt$, so that we find:
\bq \label{pdot1}
M(t) \,  \frac{dp}{dt}  = -\int dx \,  \,  \rho(x-q(t),t)  \, \frac{\partial V}{\partial x}.
 \eq
Again changing variable to  $y = x - q(t)$, we find
 \bq \label{mdpdt1}
M(t)  \, \frac{dp}{dt} =- \frac{\partial}{\partial q(t) } \int dy \,  \rho (y,t) \, [V(y+q(t)] . 
\eq
By dividing  Eq.  \eqref{mdot}  by Eq. \eqref{pdot1}, we find that 
\bq \label{dmdp} 
\frac{d \log M}{dp} =- 2  \frac{ \int dx \,  \, W(x) \, \rho (x-q(t),t) } {\int dx \,  \,  \rho(x-q(t) ,t)  \, \frac{ d V(x) }{d x}} .
\eq

Now suppose we have (as a result of some symmetry such as supersymmetry) that

\bq \label{susy}
W(x) = C_1 \,  \frac{dV(x) }{d x}.
\eq
Then
% \bq \label{dmdt1} 
%\frac {dM}{dt} =2 C_1 \frac{\partial}{\partial q(t) } \int dy \rho (y,t)V(y+q(t),t) 
%\eq
%From Eqs. \eqref{mdpdt1}, \eqref{dmdt1} we then obtain
\bq
\frac{d \log M}{dp}= - 2 C_1.
\eq
Integrating we obtain a conservation law
\bq \label{law1}
\log M + 2C_1 \, p = C_2.
\eq
This is quite similar to Eq. (10) of Kominis \cite{kominis} , however he has $v$ instead of $2p$ which is not correct when the potential has an imaginary part. In both approaches, the resulting conservation law  reduces the space  of these particle-like variables so it is confined to a two dimensional subspace. However the correct subspace is  in the variables  $p, M $ and not $\dot q, M$.
We can also introduce the (unnormalized) two-point correlation function, where again $y = x - q(t)$:
\bq
G_2(t) = \int dy ~y^2  \, \rho(y,t).
\eq
Multiplying the continuity equation by $x^2$, integrating over all space and then changing variables to $y$ one finds:
\bq
\frac{dG_2}{dt } =2  \int dy ~y~ j(y,t)+2  \int dy ~y ~\rho(y,t) \, W[y+q] .
\eq
\section{Collective Coordinate Approach to solitary wave behavior in complex  potentials}

We start with the exact solution  \cite{drazin} for the solitary wave in the NLSE when the  potential is zero for the case $\kappa=1, g=2$, namely:
\bq
\psi(x,t) = \beta \,   \sech [ \beta (x-vt)] e^{i  \left[p (x-vt) - \phi(t) \right]},
\eq
where
\bq \label{exact2}
p= \frac{v}{2}; ~~ \phi(t) = -\left( \frac{v^2}{4} + \beta^2 \right) t + \phi_0 .
\eq
The mass $M$ of the solitary wave is defined as 
\bq
M = \int  dx \,  \pstar \psi = \beta \int  dy \,  \sech^2 y = 2 \beta,
\eq
and the momentum $P$ is defined as
\bq
P(t) =  \frac{1}{2}  \int dx \,  j(x) =  M(t) p(t). 
\eq

We next assume that we can parametrize the  ``approximate'' solitary wave by the {\it  same}  parameters that the solitary wave has when the potential is zero, with the difference being  that  $\beta \rightarrow \beta(t)$,
$vt \rightarrow q(t)$, $p \rightarrow p(t)$, and  $\phi(t) $ now are  unspecified functions of $t$ \cite{clss}.  That is, we will take as our trial wave function:

\bq\label{trial1}
\psi(x,t) = \beta(t)  \sech [ \beta(t) (x-q(t) )] e^{i \left[p(t)  (x-q(t) ) - \phi(t) \right]}.
\eq

For the free part of the Lagrangian, we get the effective free action
\bq
S_0 = \int dt L_0 = \int dt M(t) \left[ p \dot q + \dot \phi - p^2 -  \frac{\beta^2}{3} \right] ,
\eq
where $M(t) = 2 \beta(t)$. The self-interaction contributes
\bq
S_I = \int dt L_I = \int dt M(t)\left[ \frac{2 \beta^2}{3}\right].
\eq

The real part of the  potential contributes:
\ba
S_v &&= - \int dt  \beta^2 (t) \int dx \,   V(x) \sech^2 [ \beta (x-q(t) )]  \nonumber \\
&& =  - \int dt  \beta (t) \int dz  V \left(\frac{z}{\beta} + q \right) \sech^2 z  \nonumber \\
&& \equiv - \int dt  ~ 2 \beta (t) U_{eff} (\beta, q) ,
\ea

where
\bq
U_{eff} (\beta, q) =  \frac{1}{2} \int dz  V \left(\frac{z}{\beta} + q \right) \sech^2 z  .
\eq
Thus 
\bq
\Gamma = \int  dt L_c = 2 \int dt  \beta (t)  \left[ p \dot q + \dot \phi - p^2 +  \frac{\beta^2}{3} - U_{eff} (\beta, q) \right].
\eq
For the dissipation function we obtain:
\ba
F&& =  i \int dx \,  dt W(x) \left( \psi_t \psi^\star -\psi_t^\star \psi \right) \nonumber \\
&& =-2  \int dt  \beta^2(t)  dx \,  W(x) \sech^2 \beta (x-q(t) ) \left[  \dot p (x-q(t)) - p \dot q - \dot \phi    \right ]  \nonumber \\
&&=-2  \int dt  \left(  \beta (t)   \int dz W \left(\frac{z}{\beta} + q \right) \sech^2 z  \left[   \frac{z} {\beta}\dot p - p \dot q - \dot \phi    \right ]   \right),  \nonumber \\
\ea
where we have set $ z = \beta (x-q)$. 
When the imaginary part of the potential is a negative constant  $W(x) \rightarrow  - \alpha$, then we obtain
\bq
F =  -\int dt \left[  4 \alpha  \beta(t) \left(p \dot q + \dot \phi    \right)  \right].
\eq
The equation for $\beta$ comes from:
\bq
\frac {d}{dt} \frac{\delta L}{\delta  \dot \phi}  = - \frac{\delta F}{\delta \dot \phi }.
\eq
For an arbitrary complex potential whose imaginary part is $W$ we obtain:
\bq
2 \dot \beta = 2 \beta(t) \int  dz W \left(\frac{z}{\beta} + q \right) \sech^2 z ,
\eq
which is just a restatement of Eq. \eqref{mdot}. 

Defining 
\bq
W_{eff} [\beta,q] = \frac{1}{2}\int  dz W \left(\frac{z}{\beta} + q \right) \sech^2 z ,
\eq
we can write this equation as
\bq \label{dotbeta}
\dot \beta = 2 \beta(t) W_{eff}[\beta,q] .
\eq

In the special case  $W(x) \rightarrow  - \alpha$ we obtain
\bq
2 \dot \beta = - 4 \alpha \beta.
\eq
Since $M(t) = 2 \beta$ for our variational ansatz, this is exactly the collective coordinate version of the  equation for the dissipation of $M(t)$ we discussed earlier [Eq. \eqref{dissM}].  

In general the equations for the collective coordinates are
\bq
\frac{\delta  \Gamma}{\delta Q_i} =  -\frac{\delta F} {\delta \dot Q_i} , 
\eq
where $Q_i= q,p,\phi, \beta$.

Choosing $Q_i=p$, we obtain from
\bq
\frac{\delta  \Gamma}{\delta p} = - \frac{\delta F} {\delta \dot p}
\eq

that 
\bq
 \dot q = 2 p +   \frac{1} {\beta}  \int dz W[ z/\beta + q] \,  z \, \sech^2 z,
\eq
which is the collective coordinate version of Eq. \eqref{litp}.
Defining
\bq
Y_{eff} [q, \beta] =  \int dz W[ z/\beta + q] ~ z ~ \sech^2 z ,
 \eq
 we can write this equation as
 \bq \label{dotq}
 \dot q = 2 p +   \frac{1} {\beta} Y_{eff} [q, \beta] .
\eq

From the Euler-Lagrange  equations:
\bq
- \frac{d}{dt} \frac{\delta L} {\delta \dot q}  + \frac{\delta L} {\delta q} = - \frac {\delta F}{\delta \dot q}
\eq
we obtain
\bq
- \frac{d}{dt} [2 \beta p] - 2 \beta \frac{ \partial U_{eff}}{\partial q} = - 2 \beta p \int dz W[ z/\beta + q]   \sech^2 z.
\eq
The last term we recognize as $- 2 \dot \beta p$, so we obtain the simple result:
\bq \label{dotp}
\dot p(t) = - \frac{\partial U_{eff}} {\partial q}.
\eq

Finally the equation for $\phi$ is obtained by choosing $Q_i = \beta$:
\bq
- \frac{d}{dt} \frac{\delta L} {\delta \dot \beta}  + \frac{\delta L} {\delta \beta } =- \frac {\delta F}{\delta \dot \beta}  \rightarrow \frac{\delta L} {\delta \beta } =0.
\eq
From this we obtain:
\bq \label{dotphi}
\dot \phi = p^2 - p \dot q - \beta^2 +  \frac {\partial }{\partial \beta} \left[ \beta U_{eff}(\beta,q) \right],
\eq
which reduces to the exact result of Eq. \eqref{exact2} when $V=0$.

\section{$\mathcal{PT}$ Symmetric Potentials}
Recently there has been much interest in $\mathcal{PT}$ symmetric potentials for the NLSE because they represent equal gain and loss in nonlinear optical devices.
Under $ \mathcal{P}$:  $x \rightarrow -x$ and under $ \mathcal{T}$:  $t \rightarrow -t$  and $i \rightarrow -i$,  the complex potential 
\bq
V(x) + i W(x) \rightarrow V(-x) - i W(-x) , 
\eq
so that if $V(x)$ is even and $W(x)$ is odd one has that the complex potential is $\mathcal{PT}$ symmetric. 
 Let us consider now  a particular potential considered by 
Kominis \cite{kominis}:  
\bq
V(x) = V_0 \cos k_0 x, ~~  W(x) = W_0 \sin l_0 x.
\eq
From this we can evaluate 
\bq
 U_{eff} (\beta, q) = \frac{V_0}{2} \int dz \cos[k_0 z/\beta + k_0 q] \sech^2 z.
\eq
Expanding the  $\cos(x)$ term and keeping only the even part we then have that the integral is
\bq
\cos k_0 q(t) \int dz \cos[ k_0 z/\beta] \sech^2 z =  2 \cos \left(k_0 q \right)\, K_0 \, \text{csch}\left(K_0\right).
\eq
Thus 
\bq
U_{eff} (\beta, q) = V_0  \cos \left(k_0 q \right) \, K_0 {\rm csch}K_0 , 
\eq
where we have introduced the notation 
\bq
K_0 =\frac{\pi k_0}{2\beta},  ~~L_0 =\frac{ \pi l_0}{2 \beta}.
\eq
To determine the dissipation function we need to evaluate
\bq
W_{eff}( \beta,q) = \frac{W_0}{2} \sin \left(l_0 q \right) \int dz \cos[ l_0 z/\beta] \sech^2 z =\ {W_0} \, \sin \left(l_0 q \right)\,  L_0 \,  \text{csch}\left(L_0\right),
\eq
as well as
\ba
Y_{eff}( \beta,q) &&= \int dz z W[z/\beta + q] \sech^2 z = W_0   \cos \left(l_0 q \right)  \int dz  z \sin [ l_0 z/\beta] \sech^2 z .
\ea
We have
\bq
 \int dz  \,  z \, \sin [ a z] \sech^2 z  =\pi  \text{csch}^2 \left(\frac{\pi  a}{2}\right) \left( \frac{\pi  a }{2} \cosh
   \left(\frac{\pi  a}{2}\right)-  \sinh \left(\frac{\pi  a}{2}\right)\right),
\eq
so that
\ba
Y_{eff}( \beta,q) &&= W_0   \cos \left(l_0 q \right) \left[ \pi  \text{csch}^2\left(L_0\right)\left(L_0
   \cosh \left(L_0\right) - \sinh \left(L_0\right)\right) \right].
\ea
From our general formalism of the previous section we now have from Eq. \eqref{dotq}
\bq \label{dotq2}
\dot q = 2p + \frac{1}{\beta} W_0   \cos \left(l_0 q \right) \left[ \pi  \text{csch}^2\left(L_0\right) \left(L_0
   \cosh \left(L_0\right) - \sinh\left(L_0\right)\right) \right].
 \eq
   From Eq. \eqref{dotp}  we obtain 
  \bq \label{dotp2}
  \dot p = {k_0 V_0} \sin \left(k_0 q \right)K_0 {\rm csch}K_0 
\eq
and from Eq. \eqref{dotbeta}
\bq \label{dotm2}
\dot \beta =  2 \beta  \, W_0 \sin \left(l_0 q \right)L_0 {\rm csch}L_0 .
\eq

Finally, once we obtain $p,q,\beta$ we can obtain $\phi$ from Eq. \eqref{dotphi}:

\bq \label{dotphi2}
\dot \phi = p^2 - p \dot q - \beta^2 +  (K_0 )^2 \,  V_0 \coth \left(K_0\right)
   \text{csch}\left(K_0\right) \cos \left(k_0 q \right).
\eq
Note that if we make the restriction $k_0 = l_0$  then indeed we satisfy the condition Eq. \eqref{susy}, and we obtain from Eqs. \eqref{dotp2} and \eqref{dotm2} that
\bq
\frac{d \log {2 \beta}}{dp}  =  \frac{2W_0}{k_0 V_0} =D_1 
\eq
leading to the conservation law we derived in general [Eq. \eqref{law1}] 
\bq \label{conservation}
\log {2 \beta} - D_1p =  \text {Constant}.
\eq
We can use this conservation law to simplify the analysis of the stability of the solitary wave. 
 If we let $ p \rightarrow  p_0 + \delta p, \beta \rightarrow  \beta_0 + \delta \beta $, and assume the variation is small,  we obtain the relation:
 \bq
\delta \beta  = D_1 \beta_0 \delta p .
 \eq 
 For the results  shown in Fig. \ref{newtrapa} we let $D_1 = 1$ so that  $\beta_0 = 1/2 $ and the amplitude of oscillations is small.  For that case, the relation that 
$\delta \beta  = \frac{1}{2}  \delta p $ is borne out in the simulations, 
showing that the simulations preserve the conservation law Eq. \eqref{law1}. 
 
 We now use Eq. \eqref{conservation}  and  let $\beta = \beta_0+D_1 \, \beta_0 \,  \delta p , \, q = q_0 +\delta q,  \,  p = p_0 + \delta p.$ Using these relations we can study the two coupled equations for 
 $\delta \dot p$ and $\delta \dot q$, which can be written as a  simple matrix equation:
 \bq \label{matrix}
Y = AX ,
\eq 
where Y is the column vector ($\delta \dot q \,  ,\delta \dot p$) and X is the column vector ($\delta q, \, \delta p$). The eigenvalues of $A$ determine the frequencies of oscillation for small oscillations of these variables.
\section{Numerical approach for solving the NLSE in the presence of
  complex potentials}

We have numerically solved Eq. (\ref{NLSE}) with the initial condition
(\ref{trial1}) using the Crank-Nicolson scheme \cite{numericalR}. We have
considered the evolution of solitary waves up to $10^3$ time units
with step size $\Delta t=10^{-3}$. The complex solitary wave in the
spatial domain  was represented on a regular grid with mesh size
$\Delta x=10^{-3}$, and   free boundary conditions were imposed. We found that during the time evolution, the shape of the mass  density  $\rho(x,t)$  was well parametrized by the form:
\bq
\rho(x,t) = \beta^2(t) \sech^2[\beta(t) (x-q(t))],
\eq
with no sign of any phonon radiation in the cases that we studied. 

%%   Here we would like to briefly discuss our approach for solving equation  Eq. \eqref{NLSE}, i.e. 
%%   \bq \label{NLSEa}
%%    i \psi_{t} + \partial_x^{2}\psi +  g (\pstar\psi)^{\kappa} \psi - (V+ i W ) \psi= 0 
%%   \ee

%%%
\section{Comparison of two different collective coordinate approaches with numerical simuations}
Here we would like to present several cases that were also studied in a two collective coordinate  (2 CC)  approach  by Kominis  \cite{kominis}. 
For comparison, we will give (in our notation) the equations used by Kominis in his approach.
For the mass equation   (in our notation $M = 2 \beta$) one obtains the same first-order differential equation: 

\bq
\dot \beta = 2 \beta \,  W_0 \sin \left(l_0 q \right)\left[(L_0 \text{csch}\left(L_0\right)\right].
\eq
However, Kominis (incorrectly) identifies $ p = \dot q /2$ and as a result obtains a second-order equation for $q(t)$, namely:
 \bq
\frac{1}{2}   \ddot q = {k_0 V_0} \sin \left(k_0 q \right) 
\, K_0 {\rm csch} K_0, 
\eq
instead of our two coupled first order equations Eqs. \eqref{dotq2} \eqref{dotp2}.
He obtains a similar conservation law when $k_0=l_0$ with our $p(t)$ replaced by $\dot q /2$.

We will see below in which situations Kominis'  ~equations lead to worse agreement  when compared with the  numerical simulations of the NLSE. In all the following plots, when we plot $p(t)$ it is only for 
the 4 CC theory.  For the 2 CC theory $p = \dot q/ 2$, and $dp/dv = 1/2$, so one can  never use our stability criterion. 

\subsection{Trapped Solitary Waves}

First we consider a case when the solitary wave is trapped at the origin and where our linear stability analysis is valid.  This is achieved by 
taking as our parameters and initial conditions:
$ V_0 = -0.01; W_0 = V_0/2; k_0 = l_0 =1; q_0 = 0.1; v_0 = 0;\beta_0 = 0.5; \phi_0 = 0 $. 
%The results  for the oscillations of $v$  are shown in fig. \ref{trapped}
 The linear stability analysis discussed in the previous section, Eq. \eqref{matrix}, yields a period $T=85.6$, which
agrees with the numerical results from the  CC equations which yields $T= 85.1$ for the period of oscillation for all the CC parameters. 
For this trapped solitary wave, the 2 CC equations   lead to almost the same result for the behavior of $\dot q$ and $\beta$ as the 4 CC equations and both agree with the numerical simulations.  
%\begin{figure}%                 use [hb] only if necceccary!
%  \centering
%  \begin{tabular}{cc}
%\ & \\
%\includegraphics[width=7.0cm]{trappedv.eps}  & 
%\quad \includegraphics[width=7.0cm]{trappedb.eps}
%\end{tabular}
% \caption{Solitary Wave in the Trapped Case in the small oscillation regime - Left Panel  Oscillations of the velocity $v(t)$, \,  Right Panel -oscillations of the amplitude $\beta(t)$ .
% Comparison of the 4  CC  theory (blue)  with the  2  CC  theory (yellow) }
% \label{trapped}
%\end{figure}

\begin{figure}[h]
 \centerline{\includegraphics[scale=0.4]{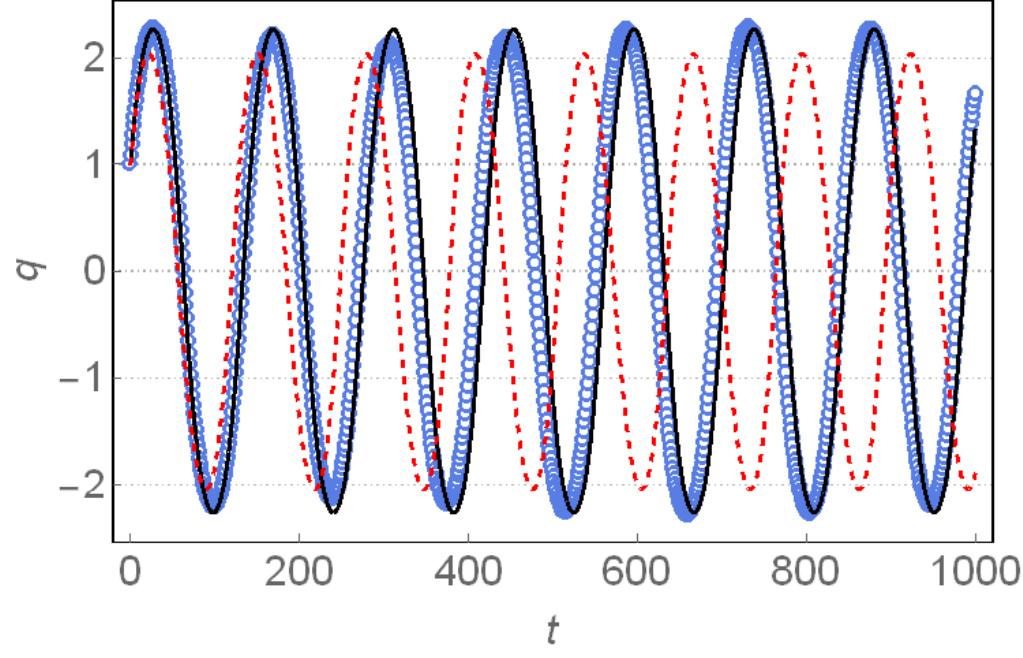}\includegraphics[scale=0.4]{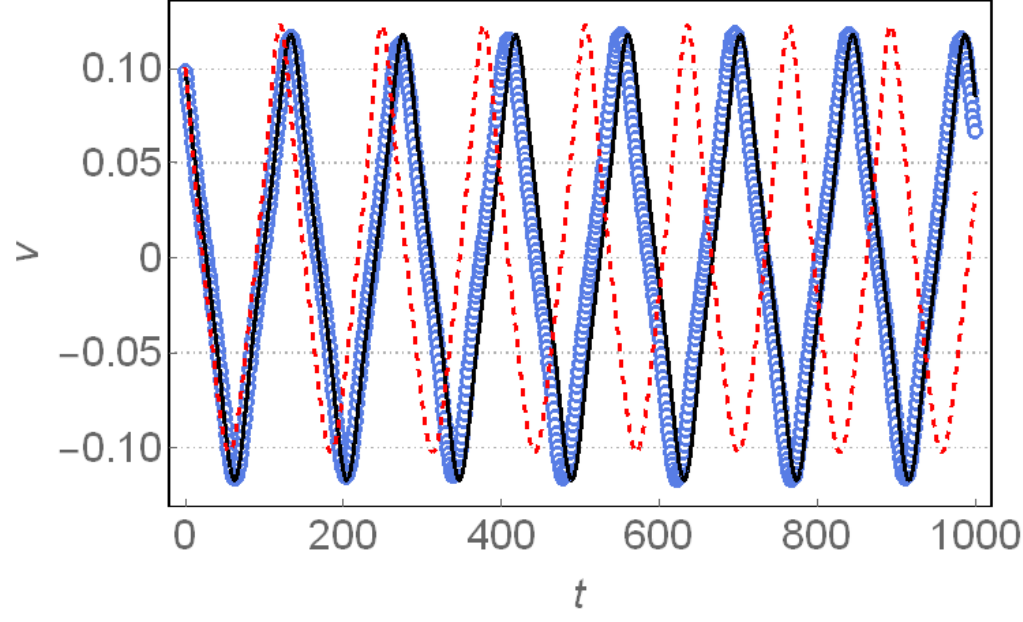}} 
 \centerline{\hspace{1cm}(a)\hspace{7cm}(b)}
 \centerline{\includegraphics[scale=0.4]{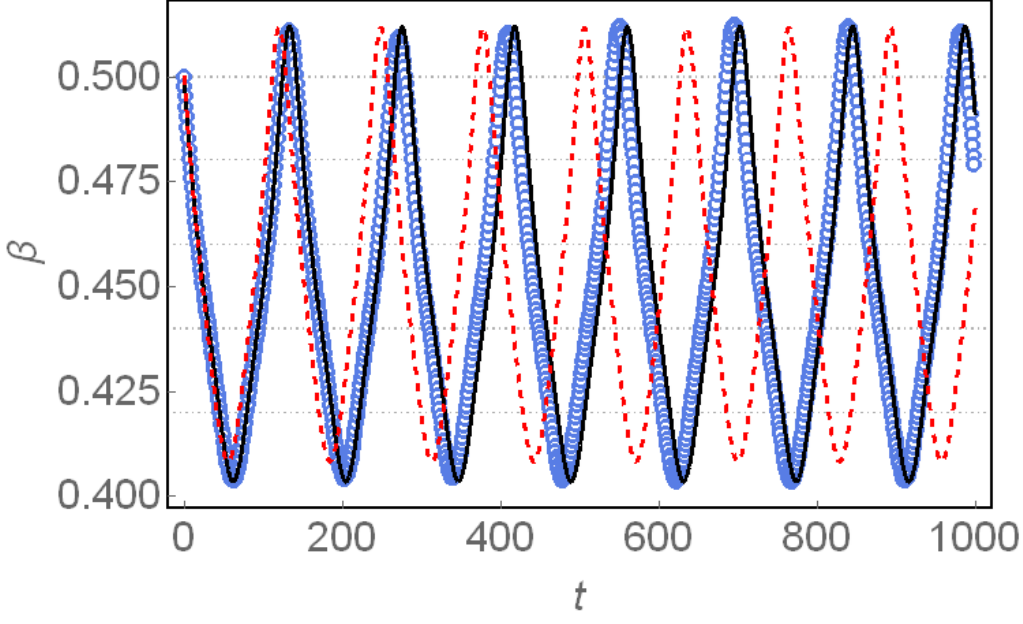}\includegraphics[scale=0.4]{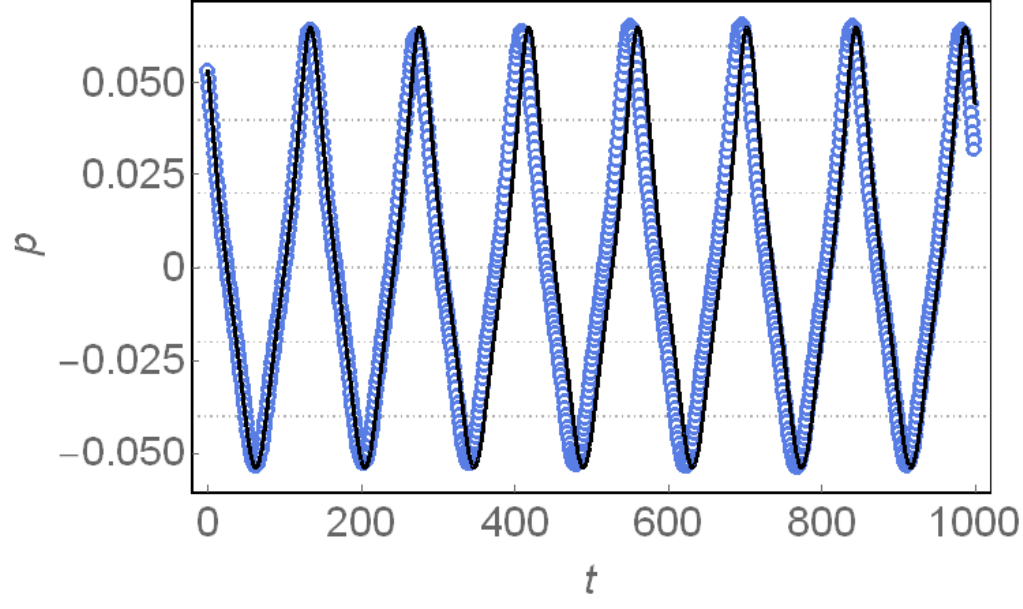}} 
 \centerline{\hspace{1cm}(c)\hspace{7cm}(d)}
 \centerline{\includegraphics[scale=0.4]{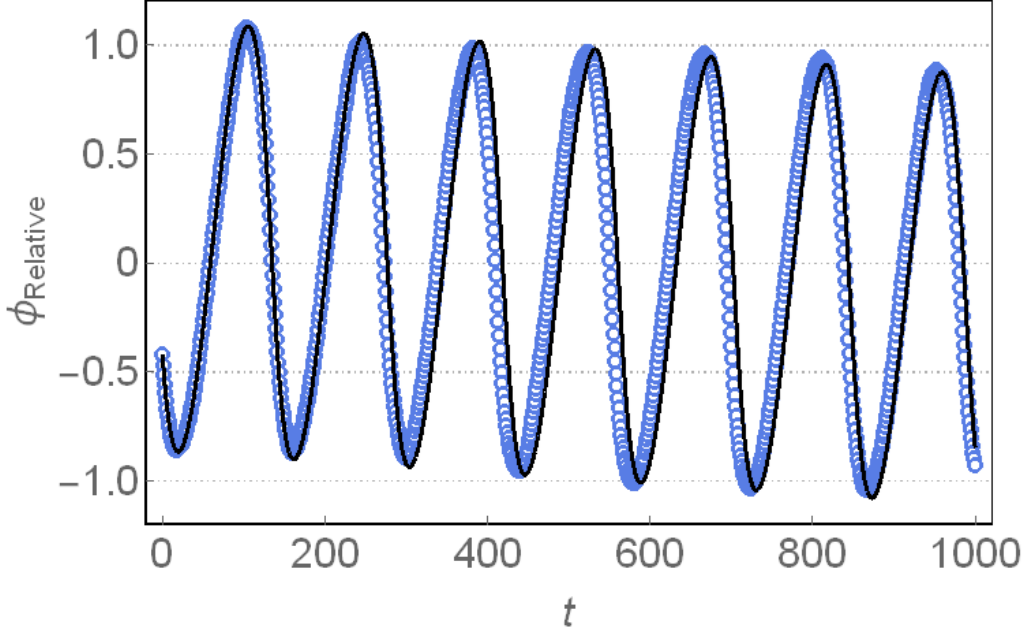}} 
 \centerline{\hspace{1cm}(e)}
  \caption{ Solitary wave in the trapped case. Comparison of the 4  CC  theory (black solid lines)  with the  2  CC  theory (red dashed lines) and numerical simulation(blue open circles) (a) position $q(t)$; (b) velocity $v(t)$; (c) amplitude $\beta(t)$; (d) momentum $p(t)$; (e)  relative phase 
  $\phi_{\textrm{relative}}(t)$. 
 $l_0=1$,   $k_0=1$, $V_0=-0.01$,   $W_0=-0.01$,  $\beta_0=0.5$, $q_0=1$, $v_0=0.1$,  $p_0=0.0531649$, 
 $\phi_0=0$.
}
 \label{newtrapa}
 \end{figure}

However, if we change the initial conditions and increase the ratio of the  strength of the imaginary to real part to be one, i.e. $W_0=V_0$ as well as change the initial
position  of the solitary wave to be one and give the solitary wave a small velocity, i.e. choose $ V_0 = -0.01; W_0 = V_0; k_0 = l_0 =1; q_0 = 1; v_0 =0.1;\beta_0 = 0.5; \phi_0 = 0, p_0= 0.0531649$,  then the  4  CC approach we advocate here gives different results than the 2 CC approach of Kominis.  In this case we are outside the range where the linear stability analysis is valid. 
A comparison of the two approximations can be seen in Fig. \ref{newtrapa}.
We find that our results (solid black curves)   agree with the numerical simulations (blue open circles) and disagree
significantly from the 2 CC equations of  Kominis (red dashed lines). 
 \noindent
 In order to compare the phases, we have subtracted the linear dependence of the phase from the original data. For that reason
  we have defined a relative phase,
 \[
 \phi_{\textrm{relative}}(t)=\phi(t)-(A\,t+B),
 \]
 where the coefficients $A$ and $B$ follow from the linear-least-squares fitting of the original data.

%\begin{figure}%                 use [hb] only if necceccary!
%  \centering
%  \begin{tabular}{cc}
%\ & \\
%\includegraphics[width=7.0cm]{trappedqa.pdf}  & 
%\quad \includegraphics[width=7.0cm]{trappedva.pdf}
%\end{tabular}
% \caption{Solitary Wave in the Trapped Case - Left Panel  Oscillations of the position $q(t)$, \,  Right Panel -oscillations of the velocity $v(t)$ .
% Comparison of the 4  CC  theory (black solid lines)  with the  2  CC  theory (red dashed lines) and numerical simulation(blue open circles)}
% \label{newtrappedq}
%\end{figure}
%
%\begin{figure}%                 use [hb] only if necceccary!
%  \centering
%  \includegraphics[width=7cm]{trappedbeta.pdf}
%  \caption{Trapped Solitary Wave -  oscillation of the amplitude $\beta(t)$. Comparison of the 4  CC  theory (black solid lines)  with the  2  CC  theory (red dashed lines) and numerical simulation(blue open circles)}  \label{newtrappedbeta}
%\end{figure}
%
Because of the conservation law, Eq. \eqref{conservation}, only the $p,q$ first order differential equations are needed.  Performing the linear stability analysis discussed earlier,
we obtain that  the period of oscillation should be $T= 118.133$ which is little  lower than that seen in the solution of the CC equations which yields $T= 142.857$.
The agreement with the linear stability analysis can be made better by decreasing the initial velocity, but this would then mask  the differences between
the outcome of using  2 CC or  4  CC  equations.

\subsection{Traveling Soliton}
\begin{figure}[h]
 \centerline{\includegraphics[scale=0.4]{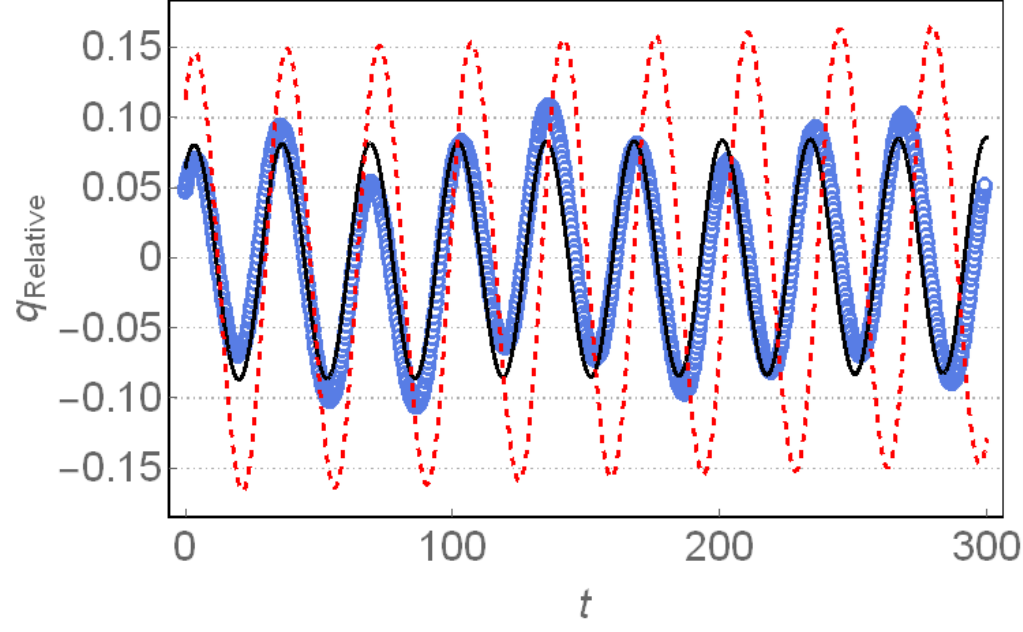}\includegraphics[scale=0.4]{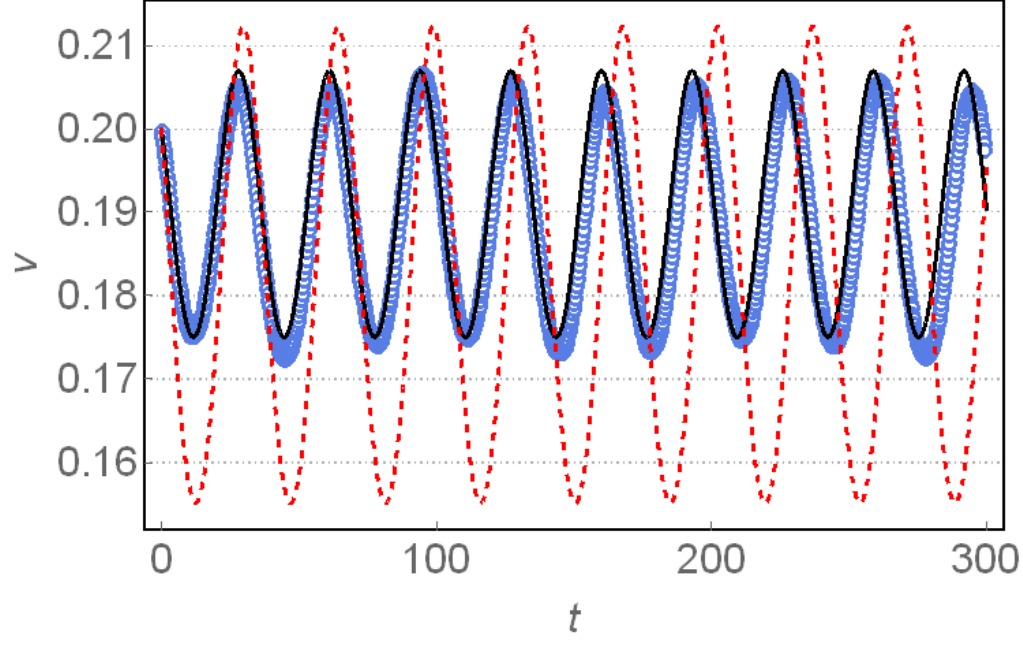}} 
 \centerline{\hspace{1cm}(a)\hspace{7cm}(b)}
 \centerline{\includegraphics[scale=0.4]{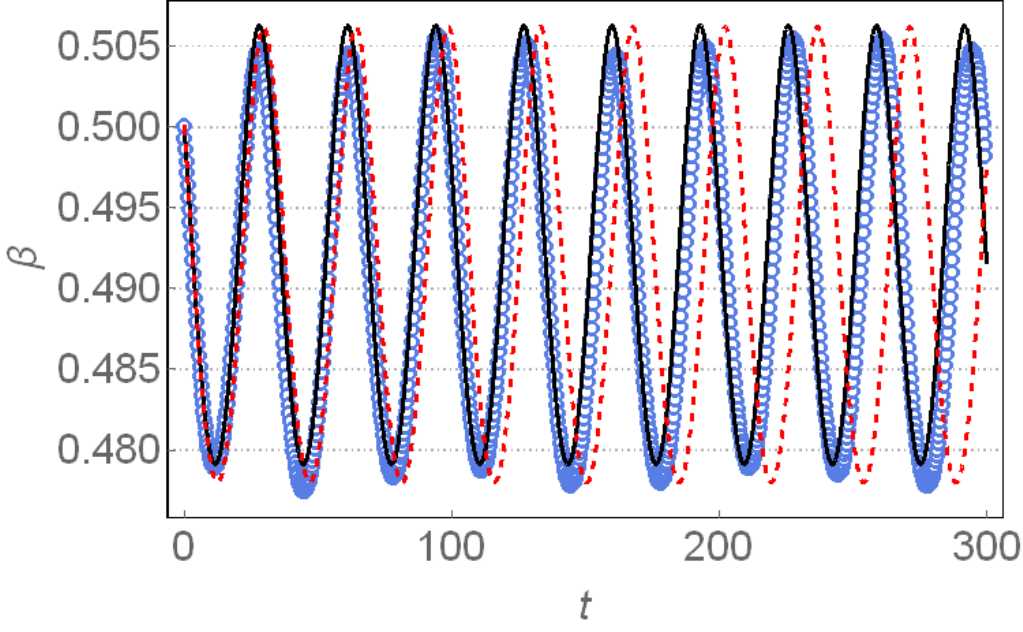}\includegraphics[scale=0.4]{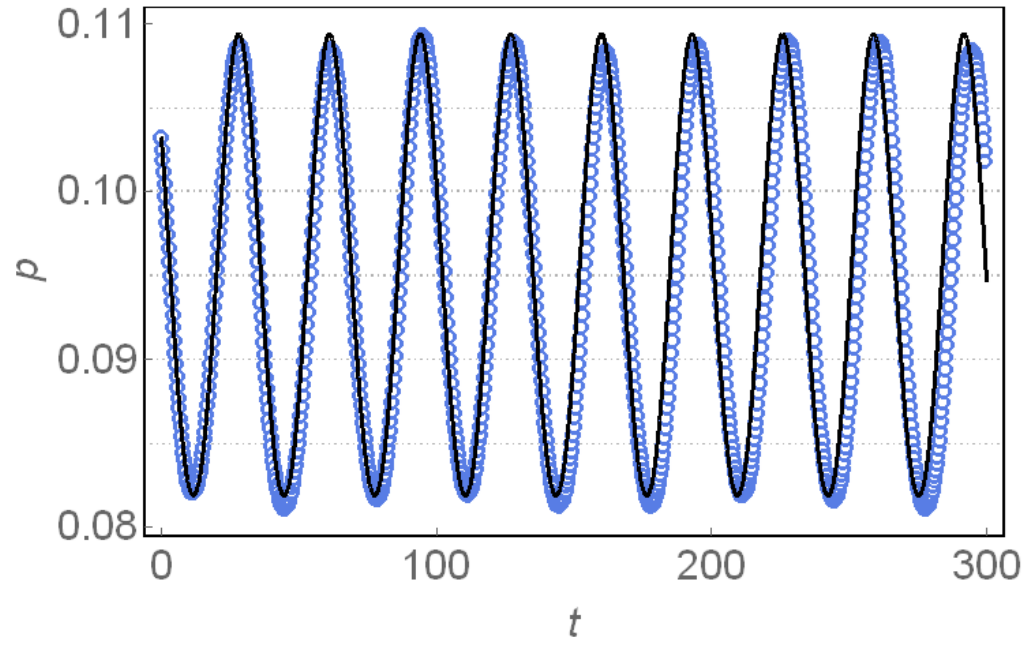}} 
 \centerline{\hspace{1cm}(c)\hspace{7cm}(d)}
 \centerline{\includegraphics[scale=0.4]{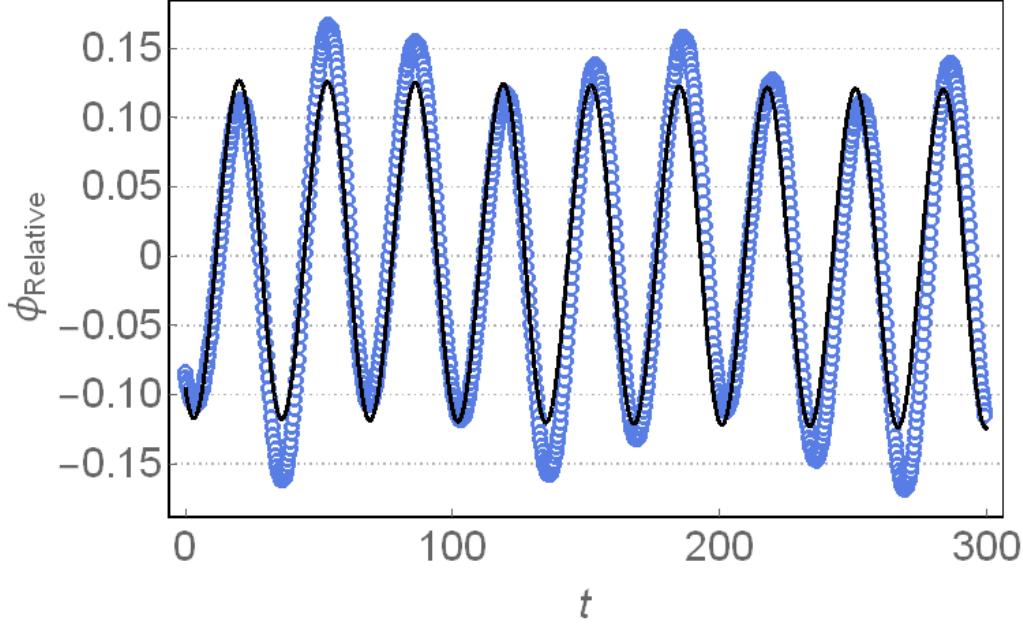}} 
 \centerline{\hspace{1cm}(e)}
  \caption{Traveling solitary wave.  Comparison of the 4  CC  theory (black solid lines)  with the  2  CC  theory (red dashed lines) and numerical simulation(blue open circles)
(a) relative position $q_{\textrm{relative}}(t)$, (b) velocity $v(t)$, (c) amplitude $\beta(t)$, (d) momentum $p(t)$, and (e) relative phase $\phi_{\textrm{relative}}(t)$.
Here  $ l_0=1$,   $k_0=1$, $V_0=−0.01$,  $ W_0=−0.01$,  $\beta_0=0.5$, $q_0=1$,  $v_0=0.2$,  $p_0=0.103165$, $\phi_0=0$.
} \label{travela} 
 \end{figure}

For the traveling soliton, one already sees instances where our 4 CC  approach differs from the 2 CC approach of Kominis.
Taking for our parameters and initial conditions:
\bq
V_0 = -0.01= W_0; k_0 = 1=l_0; \, q_0 = 1; v_0 = 0.2;  \beta_0 = 0.5; \phi_0 = 0; p_0 =.0531649;  \phi_0=0,
\eq
we are again in a situation where the conservation law  Eq. \eqref{conservation} holds.  We find  in this case that our results (black solid lines)    agree with the numerical simulations (blue open circles)  and differ significantly from the 2 CC approach of Kominis (red dashed lines) as seen in Fig. \ref{travela}.

 \noindent
 As it was done with the phase, here we have subtracted the linear dependence of the position from the original data. 
 Therefore,  we have defined a relative position,
 \[
 q_{\textrm{relative}}(t)=q(t)-(A\,t+B),
 \]
 where the coefficients $A$ and $B$ follow from the linear-least-squares fitting of the original data. 

 \noindent

%\begin{figure}%                 use [hb] only if necceccary!
%  \centering
%  \begin{tabular}{cc}
%\ & \\
%\includegraphics[width=7.0cm]{travelv.eps}  & 
%\quad \includegraphics[width=7.0cm]{travelbeta.eps}
%\end{tabular}
% \caption{Moving Solitary Wave  - Left Panel  Oscillations of the velocity $v(t)$, \,  Right Panel -oscillations of the Amplitude $\beta(t)$ .
% Comparison of the 4  CC  theory (blue)  with the  2  CC  theory (yellow) }
% \label{travel}
%\end{figure}

\subsection{Results with $k_0 \neq l_0$}
When $k_0 \neq l_0$, then the simple relation between $\beta(t) $ and $p(t)$ no longer holds and the phase space is now three dimensional. In this case our 
 4  CC  approach can differ significantly from the approach of Kominis and also we can understand when there is an instability.  First let us consider a case where the solitary wave is quasi-periodic. For this case we choose for our initial conditions:
\bq 
V_0 = -0.01, W_0 = V_0, k_0 = 1, l_0 = \sqrt{2}, 
q_0 = \pi, v_0 = 0.05;, \beta_0 = 0.5, \phi_0 = 0, p_0=0.0243222.
\eq
For this case   our 4 CC approach again agrees quite well with the numerical simulation. The 2 CC approach generally agrees with the 4 CC approach in this case but does not give information about the phase $\phi$. The quasiperiodicity is seen best in the soliton amplitude $\beta(t)$ and phase $ \phi(t)$, see Figs. 3 (c), (e). In the other CCs the quasiperiodicity is less pronounced. This difference is also obvious in the Discrete Fourier Transforms (DFT) of  $\beta(t)$ and $v(t) $ in Fig. 5.

Also  in  our approach  $p \neq {\dot q}/2$  and we have a criterion for when the period is about to change--namely when $dp/dv =0$, as shown
in Fig. \ref{quasipervp} which agrees with the numerical simulations.

 \noindent
  \begin{figure}[h]
 \centerline{\includegraphics[scale=0.4]{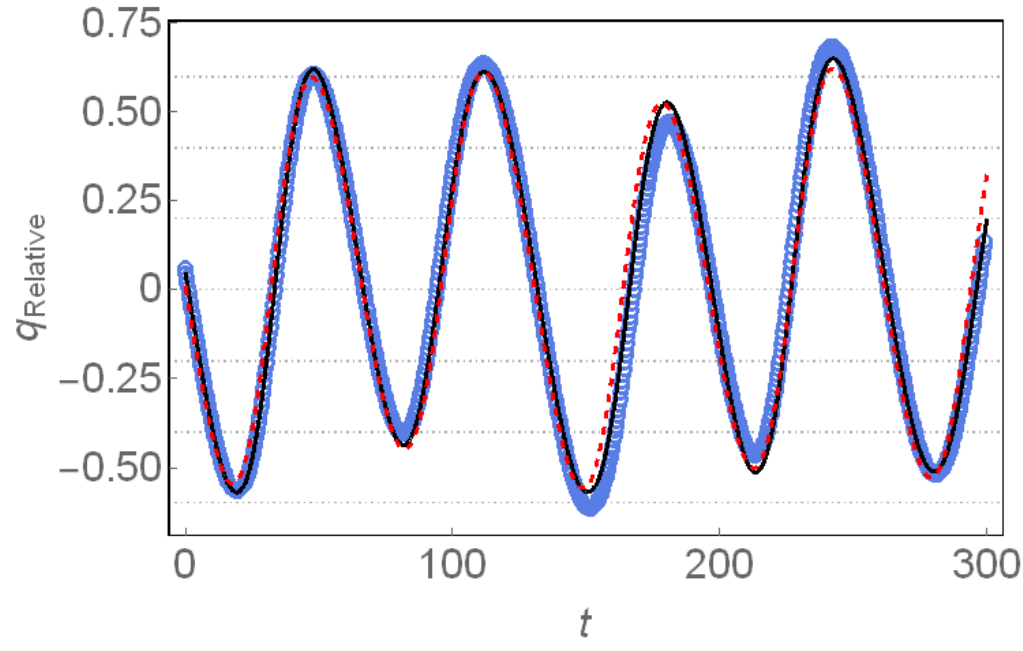}\includegraphics[scale=0.4]{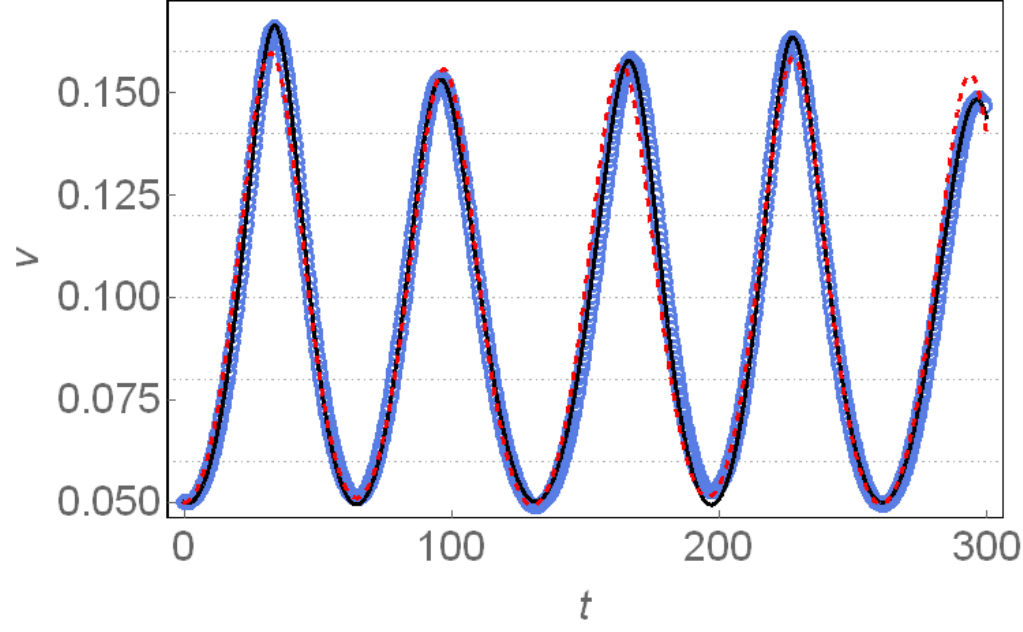}} 
 \centerline{\hspace{1cm}(a)\hspace{7cm}(b)}
 \centerline{\includegraphics[scale=0.4]{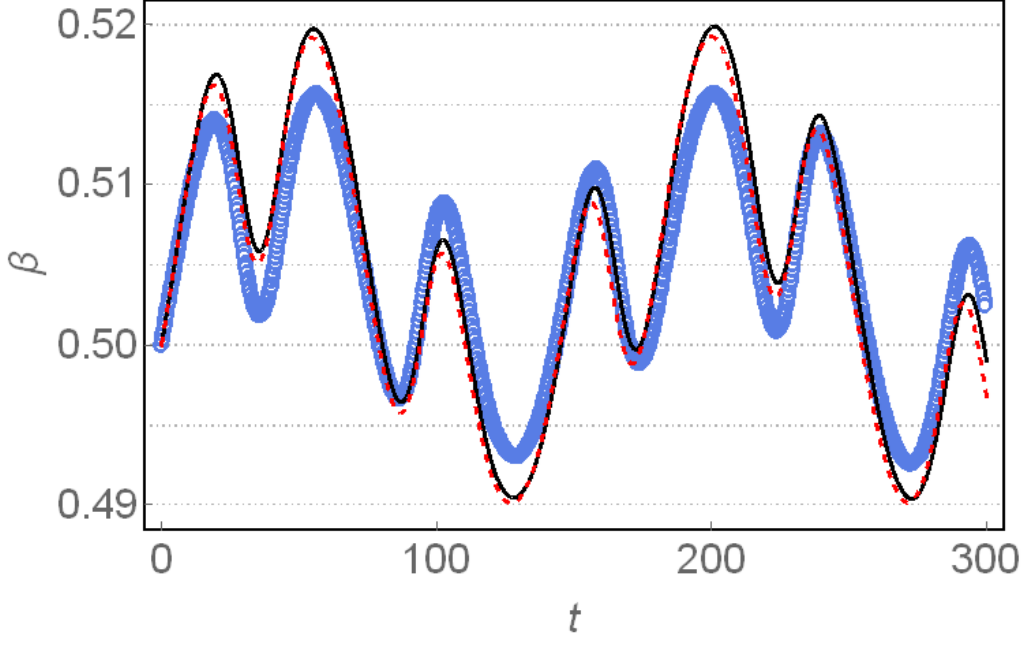}\includegraphics[scale=0.4]{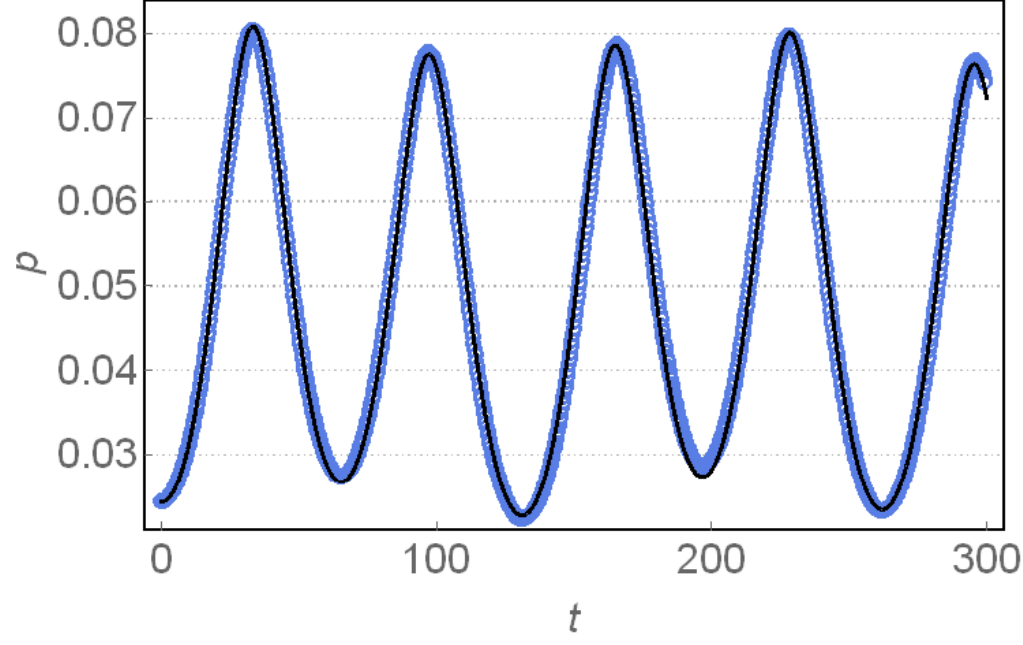}} 
 \centerline{\hspace{1cm}(c)\hspace{7cm}(d)}
 \centerline{\includegraphics[scale=0.4]{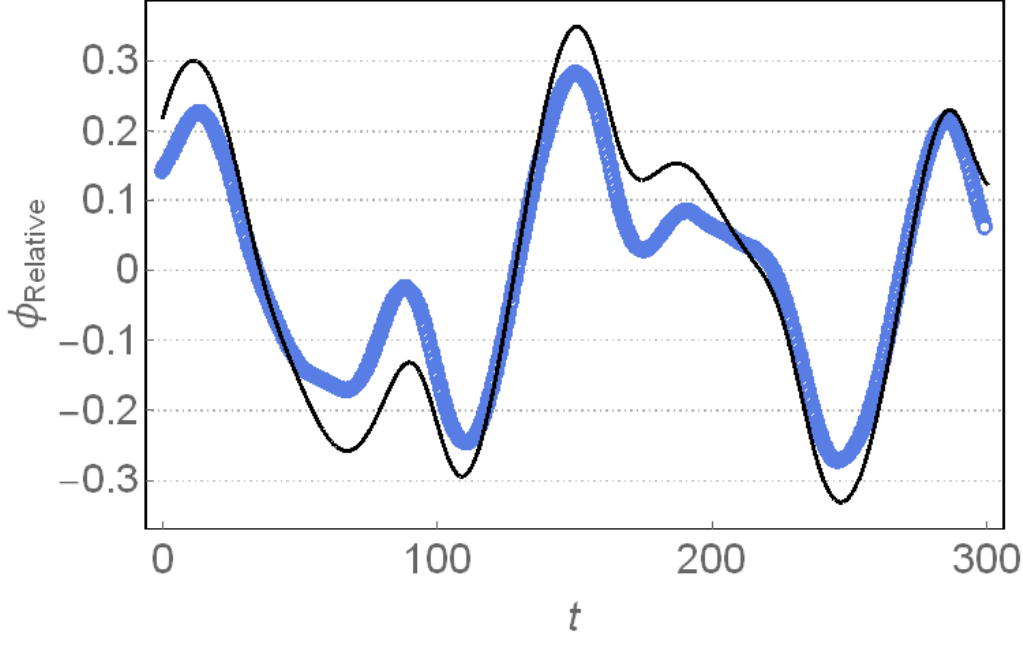}} 
 \centerline{\hspace{1cm}(e)}
 \caption{Moving solitary wave in the quasiperiodic case, 
 (a) relative position $q_{\textrm{relative}}(t)$, (b) velocity $v(t)$, (c) amplitude $\beta(t)$, (d) momentum $p(t)$, and (e)  relative phase $\phi_{\textrm{relative}}(t)$.
Here $ l_0=\sqrt{2}$,  $ k_0=1$,  $V_0=−0.01$,   $W_0=−0.01$,  $\beta_0=0.5$,  $q_0=\pi$,  $v_0=0.05$, $p_0=0.0243222$, $\phi_0=0$.
 }
 \label{quasi}
 \end{figure}

 As was done previously, here we have subtracted the linear dependence of the position as well as the phase from the original data using a linear fit.

%\begin{figure}%                 use [hb] only if necceccary!
%  \centering
%  \begin{tabular}{cc}
%\ & \\
%\includegraphics[width=7.0cm]{quasiv.eps}  & 
%\quad \includegraphics[width=7.0cm]{quasiperb.eps}
%\end{tabular}
% \caption{Moving Solitary Wave in Quasiperiodic case, $k_0=1;\, l_0=\sqrt{2}$.   - Left Panel  Oscillations of the velocity $v(t)$, \,  Right Panel -oscillations of the Amplitude $\beta(t)$ .}
% \label{quasi}
%\end{figure}

% Comparison of the 4  CC  theory (blue)  with the  2  CC  theory (yellow) }
% \label{quasi}
%\end{figure}
%
%\begin{figure}
% \centering
%  \includegraphics[width=10cm]{quasiv.eps}
%  \caption{Moving Solitary Wave in Quasiperiodic Case  velocity oscillations- $v(t)$.  Blue curve is our 4  CC  approximation, yellow curve is the 2  CC  method of Kominis}
%  \label{quasiv}
%\end{figure}
%
%
%\begin{figure}
%
%\centering
% \includegraphics[width=10cm]{quasiperb.eps}
%   \caption{Moving Solitary Wave in Quasiperiodic Case - width (Mass)   oscillation $\beta (t)$ . Blue curve is our 4  CC  approximation, yellow curve is the 2  CC  method of Kominis}
% \label{quasib}
%\end{figure}

\begin{figure}
\centering
 \includegraphics[width=7cm]{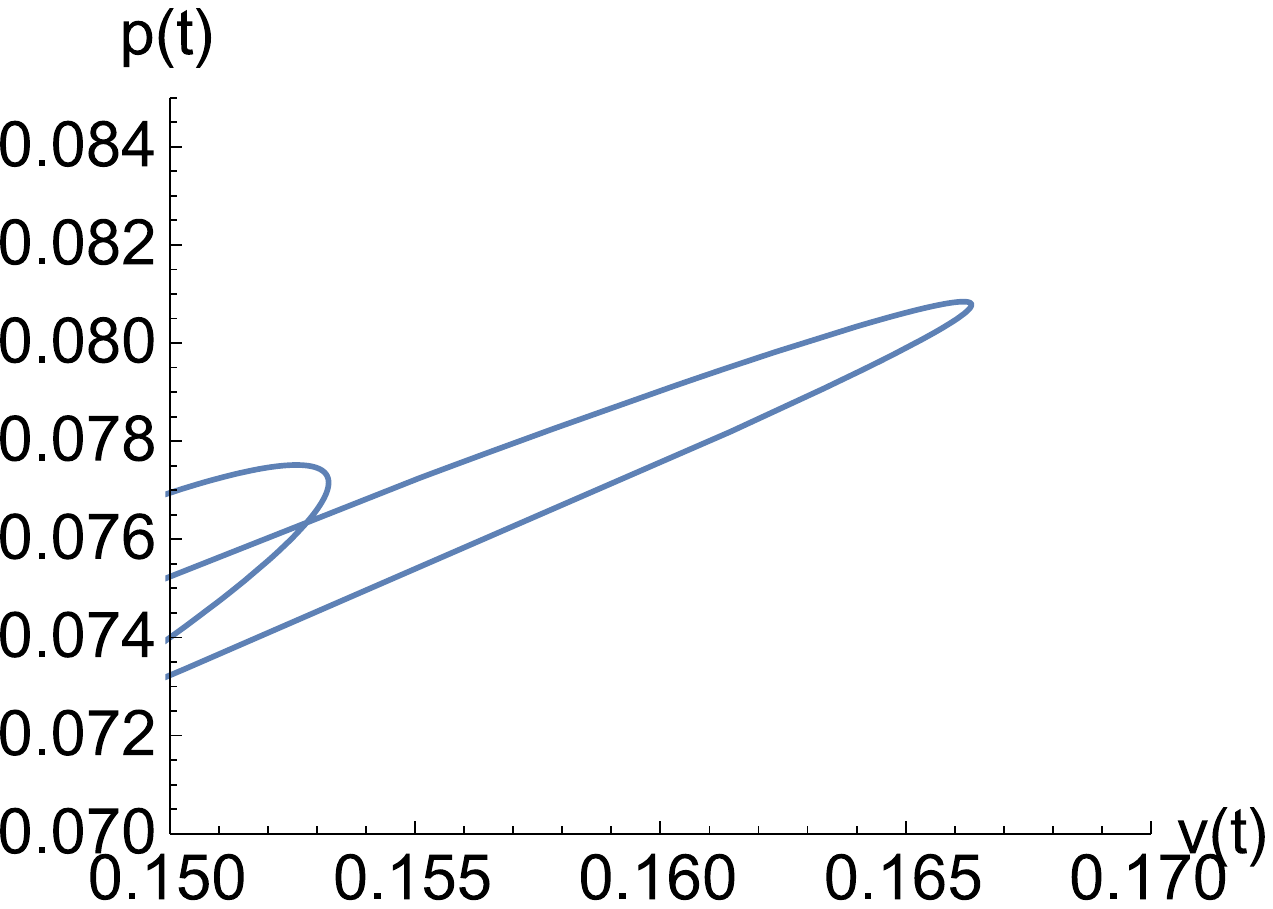}
   \caption{Moving solitary wave in the quasiperiodic case when  $k_0=1, \, l_0=\sqrt{2}$.  Magnification of the turnaround in the parametric plot of the momentum $p(t)$ vs. velocity $v(t)$.  The
   slope $dp/dv$ is negative for short pieces of the curve.}  \label{quasipervp}
\end{figure}

%Also in the quasiperiodic case the Discrete Fourier Transforms (DFT)  of $v(t)$ and $\beta(t)$ are different which is quite unusual. This is seen in  fig. \ref{dft1}.

 \begin{figure}[ht!] 
 \label{dft1}
\centering
\begin{tabular}{cc}
\ & \\
\includegraphics[width=7.0cm]{{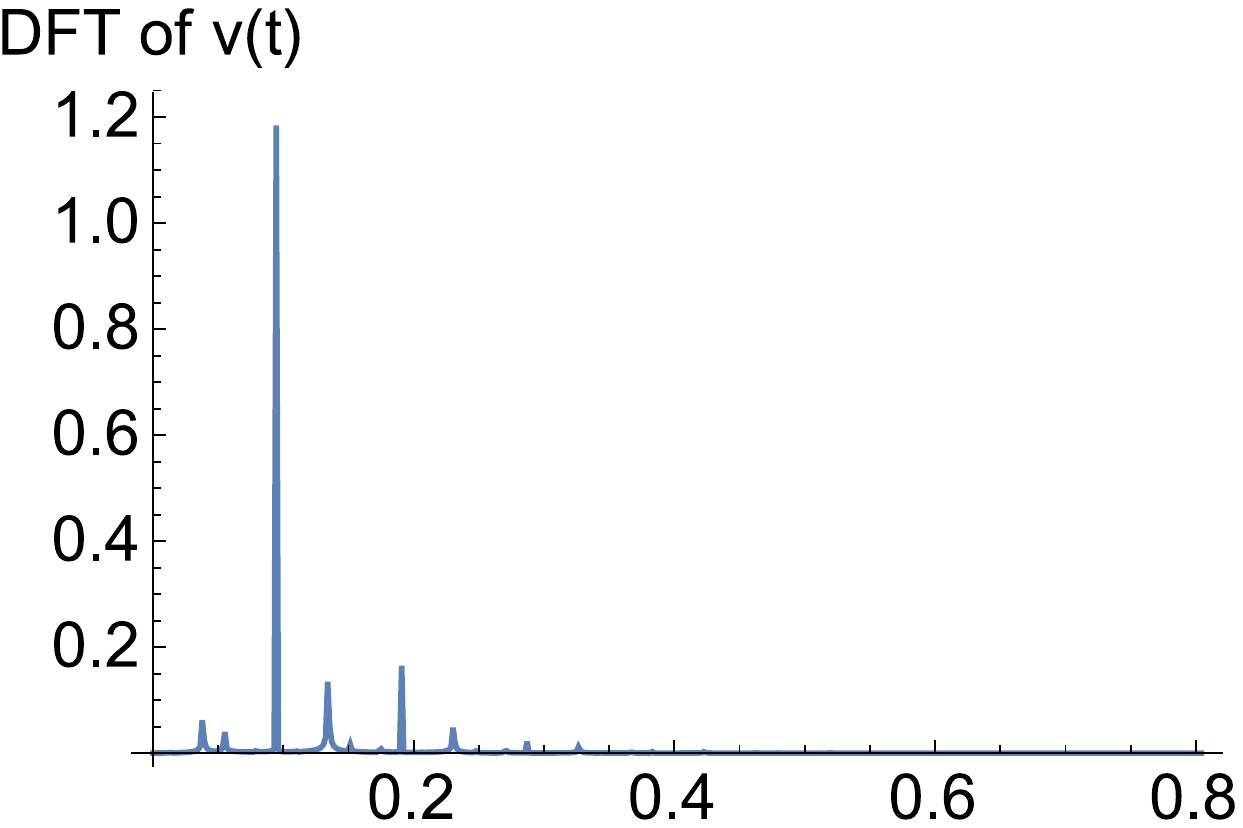}}  & 
\quad \includegraphics[width=7.0cm]{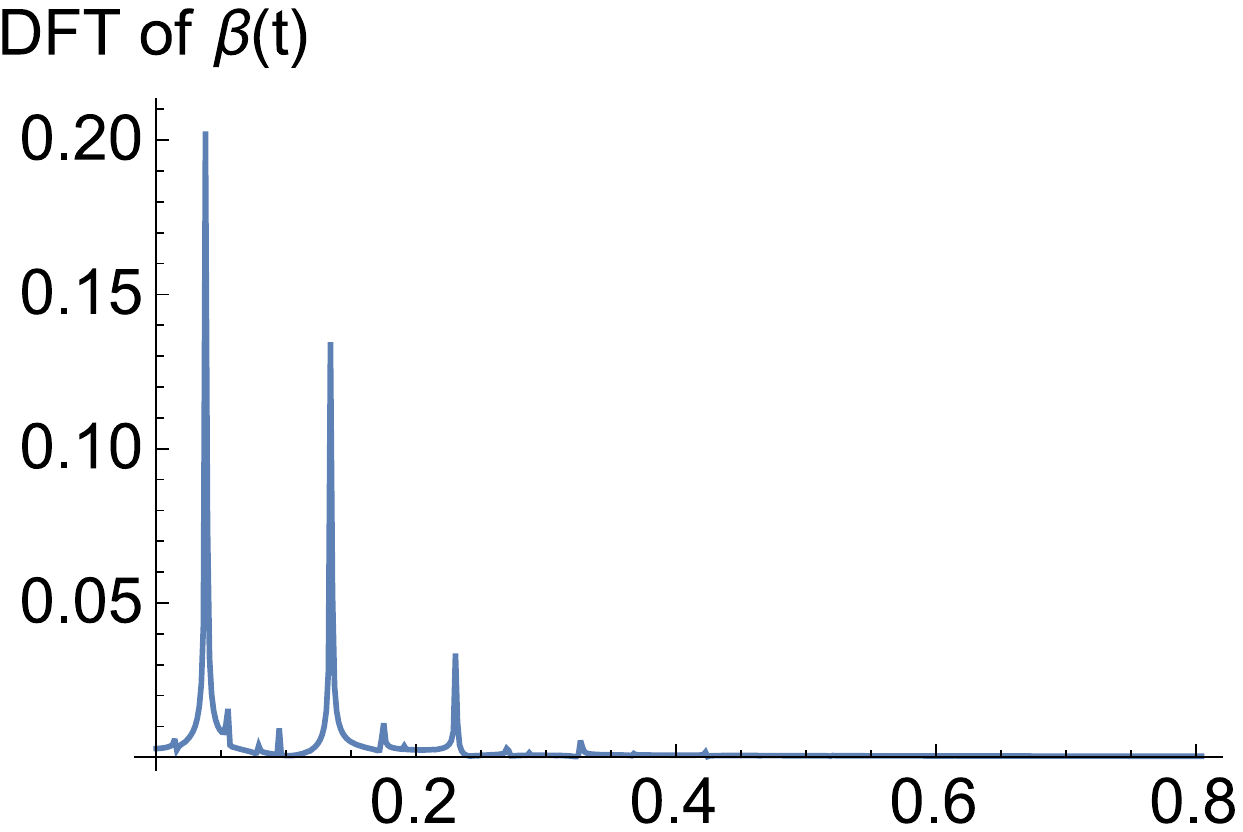}
\end{tabular}
 \caption{Moving solitary wave in the  quasiperiodic case. Left Panel: Discrete Fourier Transform of $v(t)$. Right Panel: DFT of $\beta(t).$}
 \label{dft1}
  \end{figure}

To explore a blowup case (amplitude increasing in time)  we will choose as our parameters:

\bq
V_0 = -0.01, W_0 = V_0, k_0 = 1, l_0 = 1/3,
q_0 = \pi, v_0 = -0.1, \beta_0 = 0.5, \phi_0 = 0.
\eq

\begin{figure}[ht!] 
 \centerline{\includegraphics[scale=0.4]{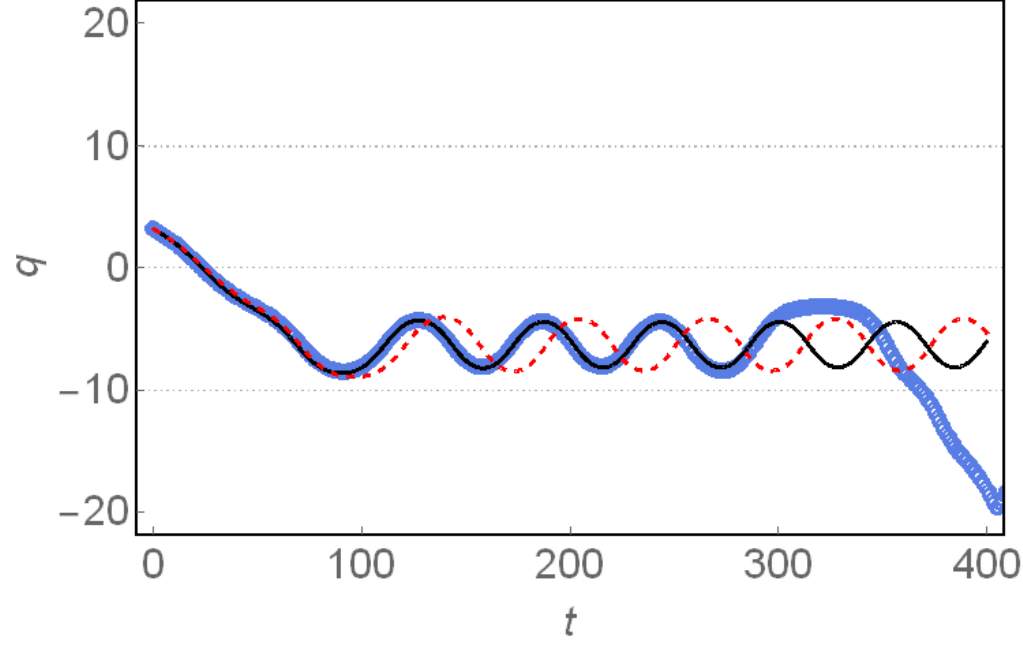}\includegraphics[scale=0.4]{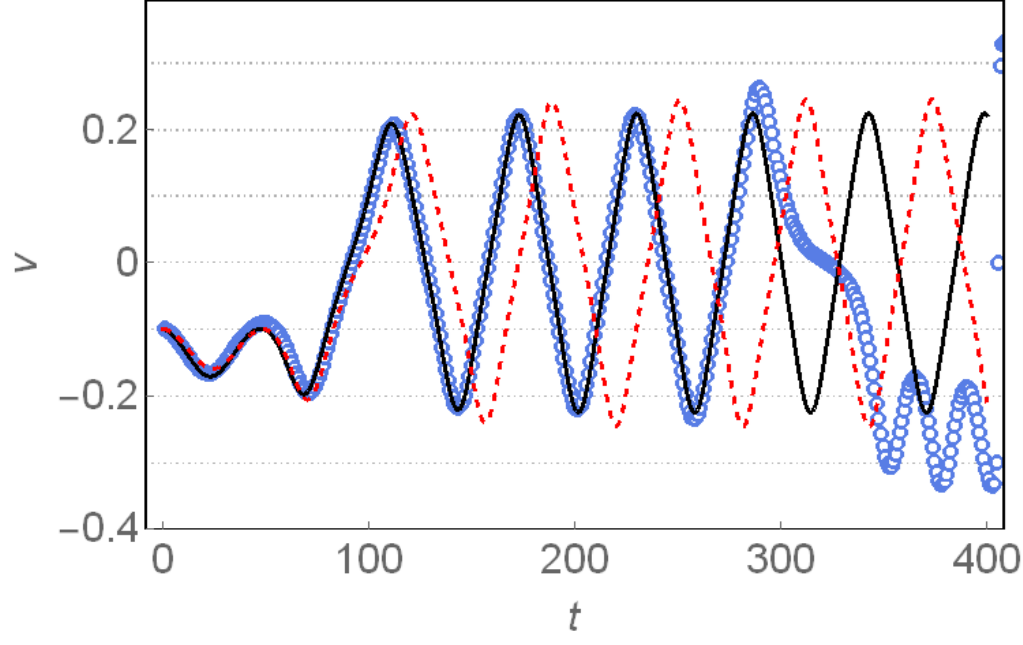}} 
 \centerline{\hspace{1cm}(a)\hspace{7cm}(b)}
 \centerline{\includegraphics[scale=0.4]{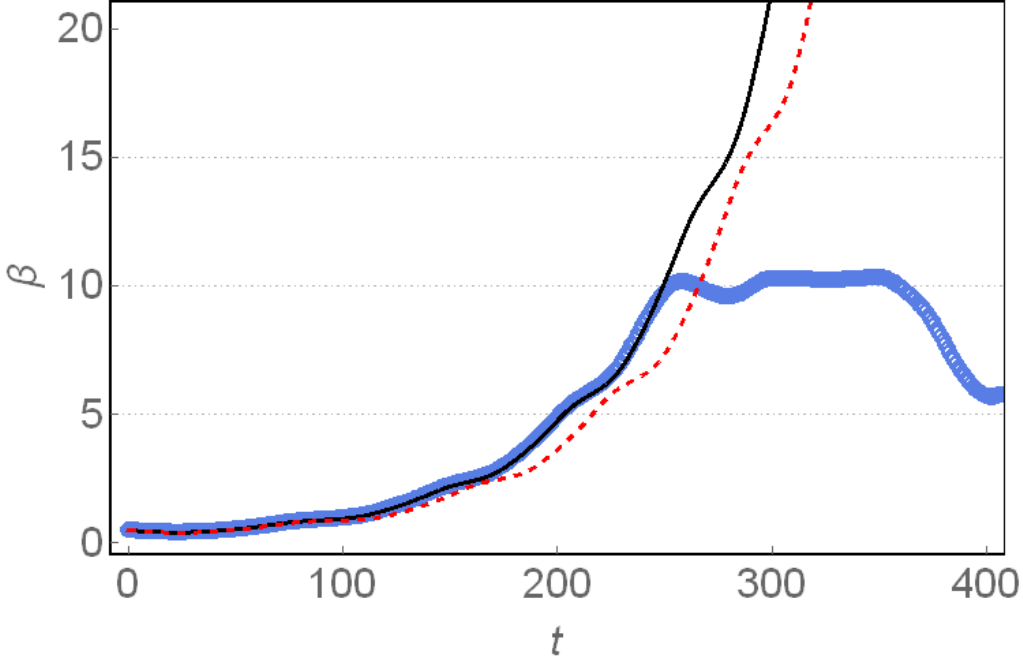}\includegraphics[scale=0.4]{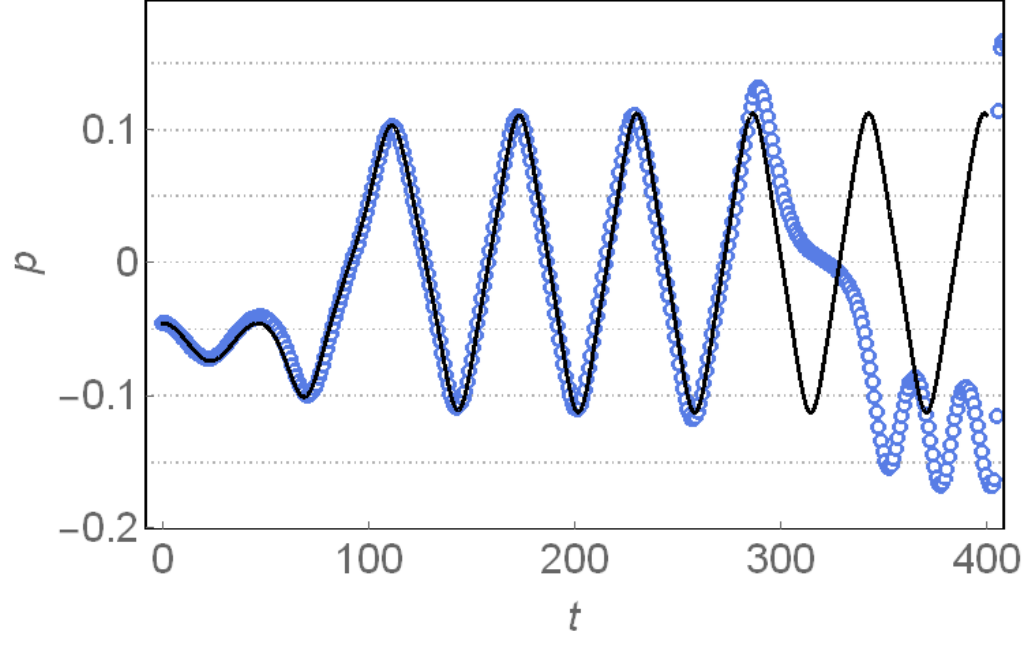}} 
 \centerline{\hspace{1cm}(c)\hspace{7cm}(d)}
 \centerline{\includegraphics[scale=0.4]{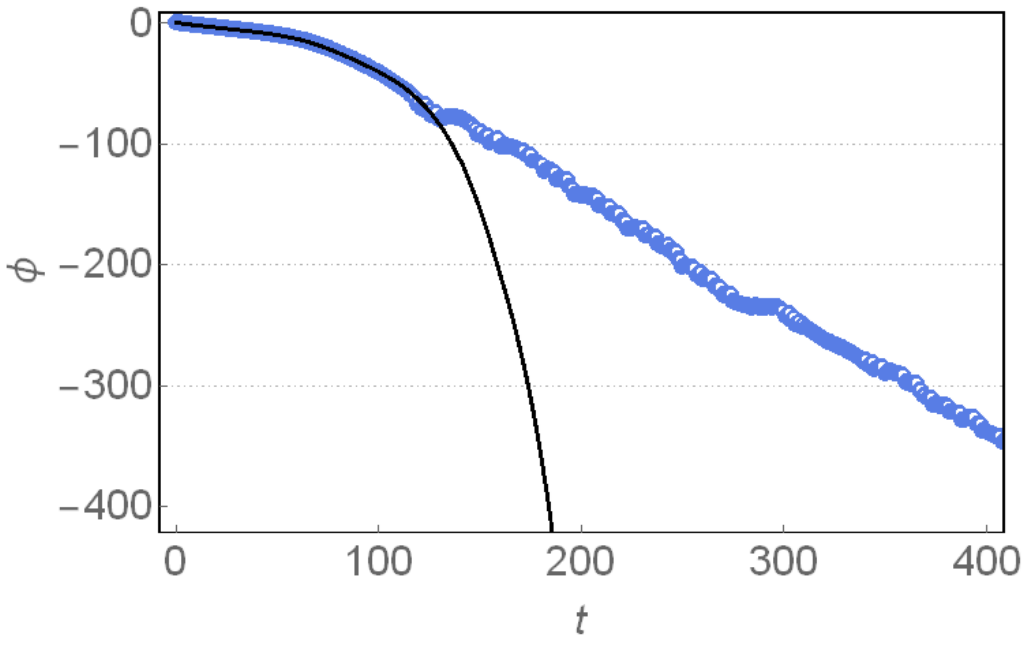}  \includegraphics[width=7.0cm]{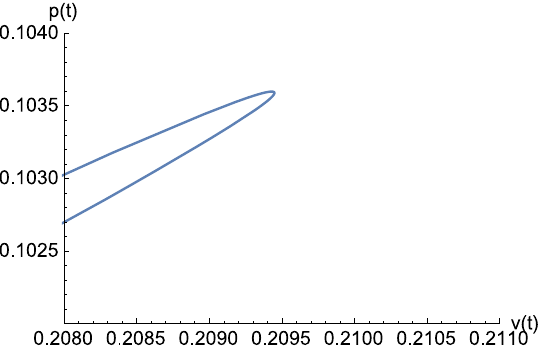}}                                                       
 \centerline{\hspace{1cm}(e) \hspace{7cm}(f)}
 \caption{ Blowup (increasing amplitude)  situation (a) position $q(t)$, (b) velocity $v(t)$, (c) amplitude $\beta(t)$, (d) momentum $p(t)$, (e) phase $\phi(t)$, and (f) magnification of the turnaround in the  parametric plot of the momentum $p(t)$ vs. velocity $v(t)$.  Parameters and initial conditions: 
  $l_0=1/3$,  $ k_0=1$,  $V_0=-0.01$,  $W_0=-0.01$, $\beta_0=0.5$,  $q_0=\pi$,  $v_0=-0.1$,  $p_0=-0.0457085$, $\phi_0=0$.
  }
  \label{blowup}
 \end{figure}

The numerical results track the 4 CC approximation up to $t \approx 300$ when the instability sets in as seen in Fig. \ref{blowup}.
Again the 2 CC approximation breaks down much earlier around $t = 100$. 

%\begin{figure}[ht!]                 %use [hb] only if necceccary!
%  \centering
%  \begin{tabular}{cc}
%\ & \\
%\includegraphics[width=7.0cm]{blowbeta.pdf}  & 
%\quad \includegraphics[width=7.0cm]{blowstab2.pdf}
%\end{tabular}
% \caption{Moving solitary wave in blowup case when $k_0=1;\, l_0=1/3 $.   - Left Panel : increase of the amplitude $\beta(t)$  \,  Right Panel : magnification of the turn around in the  parametric plot of the momentum $p(t)$ vs.  the velocity $v (t)$.}
% \label{blow3}
%\end{figure}

%\begin{figure}
%\centering
% \includegraphics[width=10cm]{blowbeta.eps}
%   \caption{Moving Solitary Wave in blowup Case -  Amplitude $\beta (t)$ . Blue curve is our 4  CC  approximation, yellow curve is the 2  CC  method of Kominis}
% \label{blowb}
%\end{figure}
%
%\begin{figure}
%\centering
% \includegraphics[width=10cm]{blowstab.eps}
%   \caption{Moving Solitary Wave in the  blowup Case -  parametric plot of the momentum $p(t)$ vs.  the velocity $v (t)$}
%     \label{stab}
%\end{figure}
We also want to relate the onset of the instabilities  to the situation when $dp/dv < 0$.   This quantity changes sign initially when $v=-0.2$ and
next when $v=0.2$ which correlates to two changes in the oscillation frequency of $v(t)$  and ultimately to the blowup of the amplitude  $\beta$ as
seen in Fig. \ref{blowup} (f).
\newpage

\subsection{Shifted Potential: ~~ $V(x) = V_0 \cos( k_0 x + \Delta)$,  $W(x) = W_0 sin l_0 x$}
We next turn to two other situations discussed by Kominis to compare the 2 CC  and  4  CC methods.  Here we shift the real part of the potential $V(x)$ away from the origin, keeping $W(x)$  unshifted.  This again breaks the conservation law. 
First we consider a case where the solitary wave is trapped but the amplitude is decreasing.  Here we choose
\bq
V_0 = -0.01, W_0 = V_0/2, K_0 = 1, L_0 = 1,  \Delta = -\pi /3, 
q_0 = 1.3 \pi /3, v_0 = 0, \beta_0 = 0.5, \phi_0 = 0.
\eq

\begin{figure}[h] 
 \centerline{\includegraphics[scale=0.4]{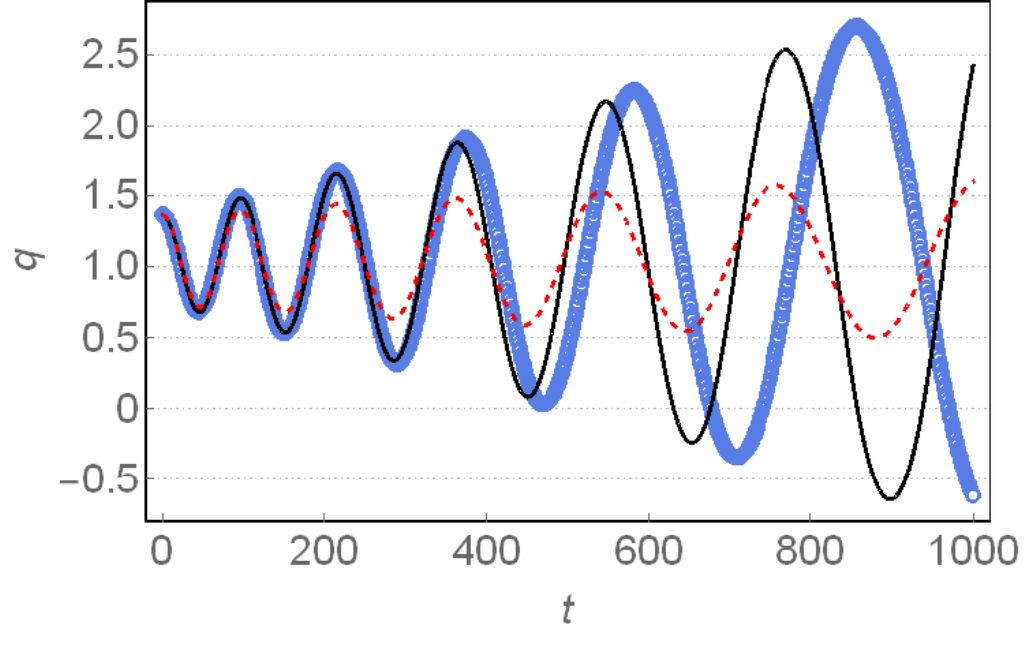}\includegraphics[scale=0.4]{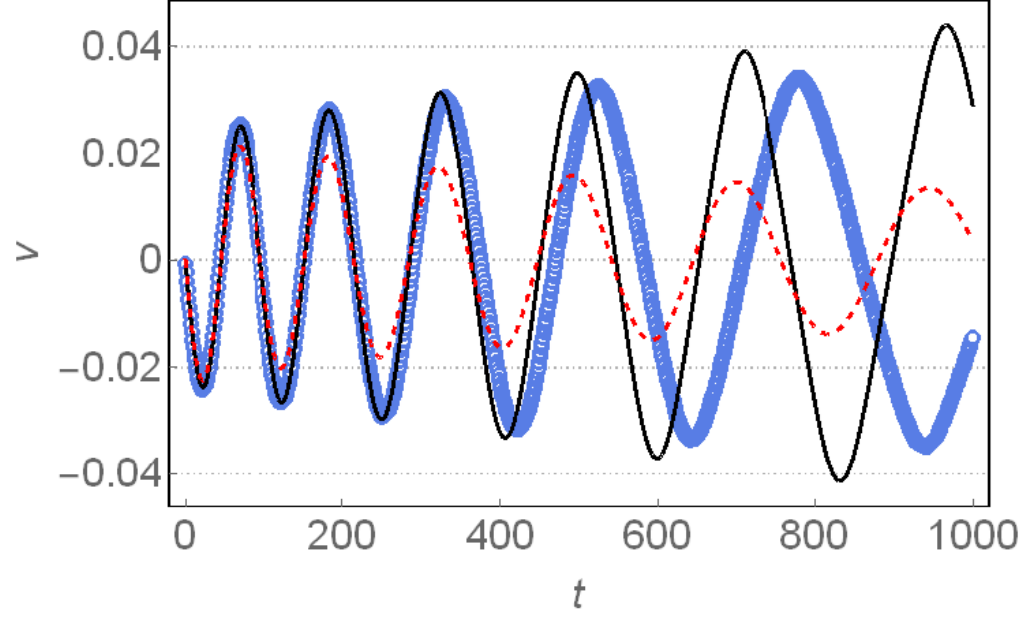}} 
 \centerline{\hspace{1cm}(a)\hspace{7cm}(b)}
 \centerline{\includegraphics[scale=0.4]{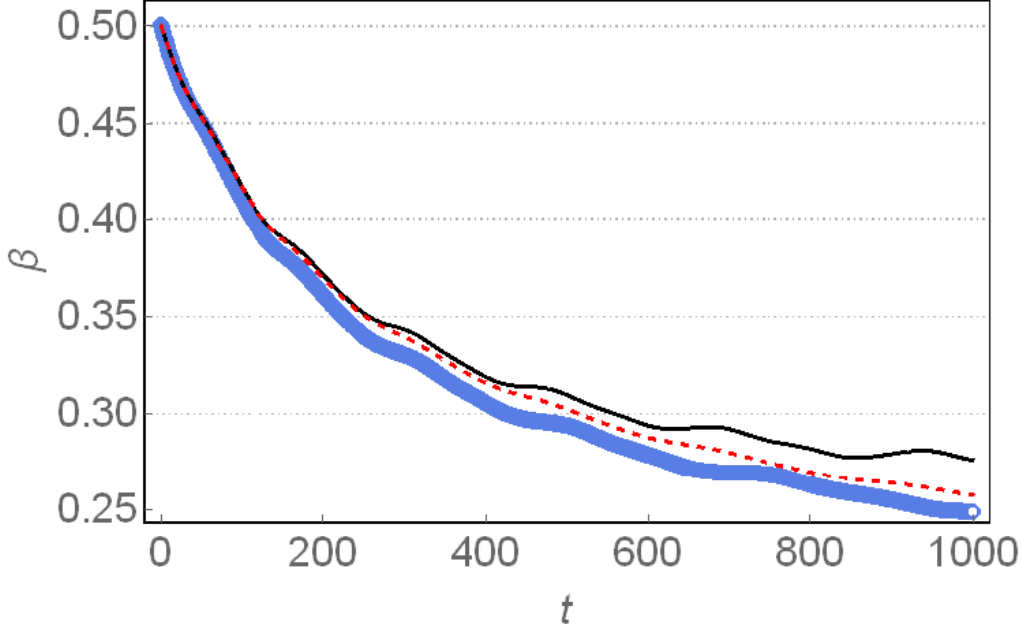}\includegraphics[scale=0.4]{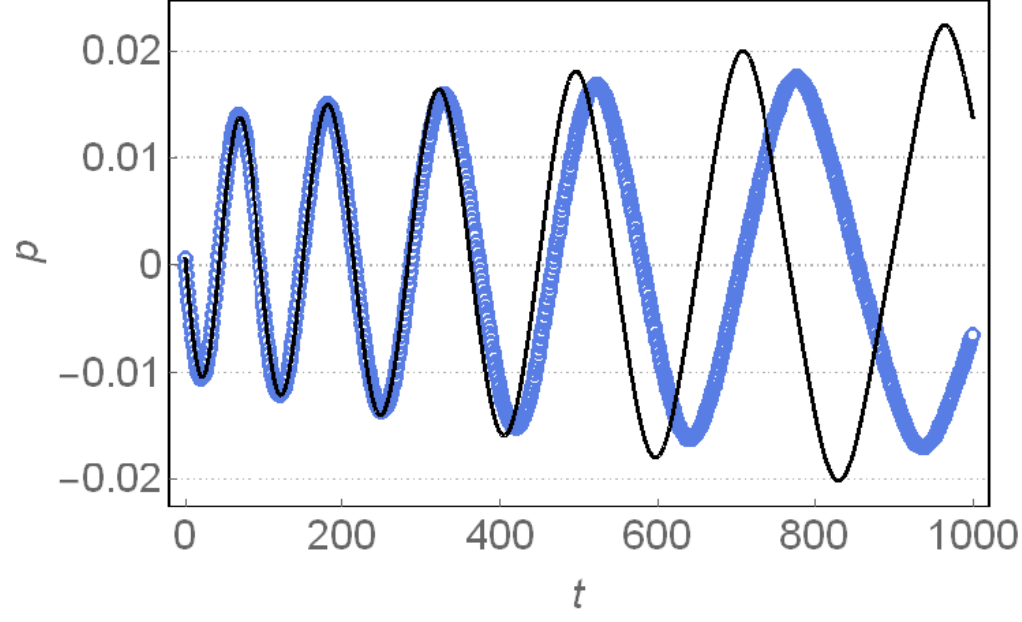}} 
 \centerline{\hspace{1cm}(c)\hspace{7cm}(d)}
 \centerline{\includegraphics[scale=0.4]{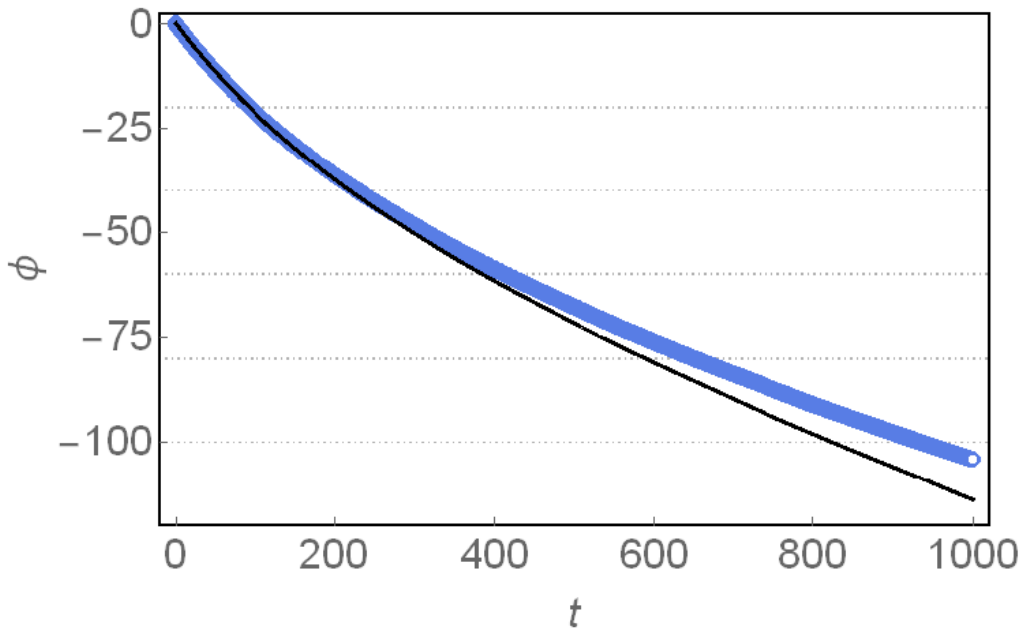}} 
 \centerline{\hspace{1cm}(e)}
  \caption{ Shifted potential-trapped case: (a) position $q(t)$, (b) velocity $v(t)$, (c) amplitude $\beta(t)$, (d) momentum $p(t)$, and (e) phase $\phi(t)$.
  Here $l_0=1$,   $k_0=1$,   $V_0=-0.01$,   $W_0=-0.005$,  $\beta_0=0.5$,  $q_0=1.3\pi/3$, $v_0=0$, $p_0=0.000608946$, $\phi_0=0$, $\Delta= -\pi/3$.
  } \label{trap1}
  \end{figure}

In this case,  the oscillations of $q(t)$, $v(t)$, and $p(t)$ increase, but the amplitude (mass) $\beta(t)$ decreases. This is seen in Fig.  \ref{trap1}.
Here we find that the 4 CC result tracks well the numerics up to $t=500$, whereas the 2 CC result begins failing around $t=200$. 

\begin{figure}
\centering
 \includegraphics[width=7cm]{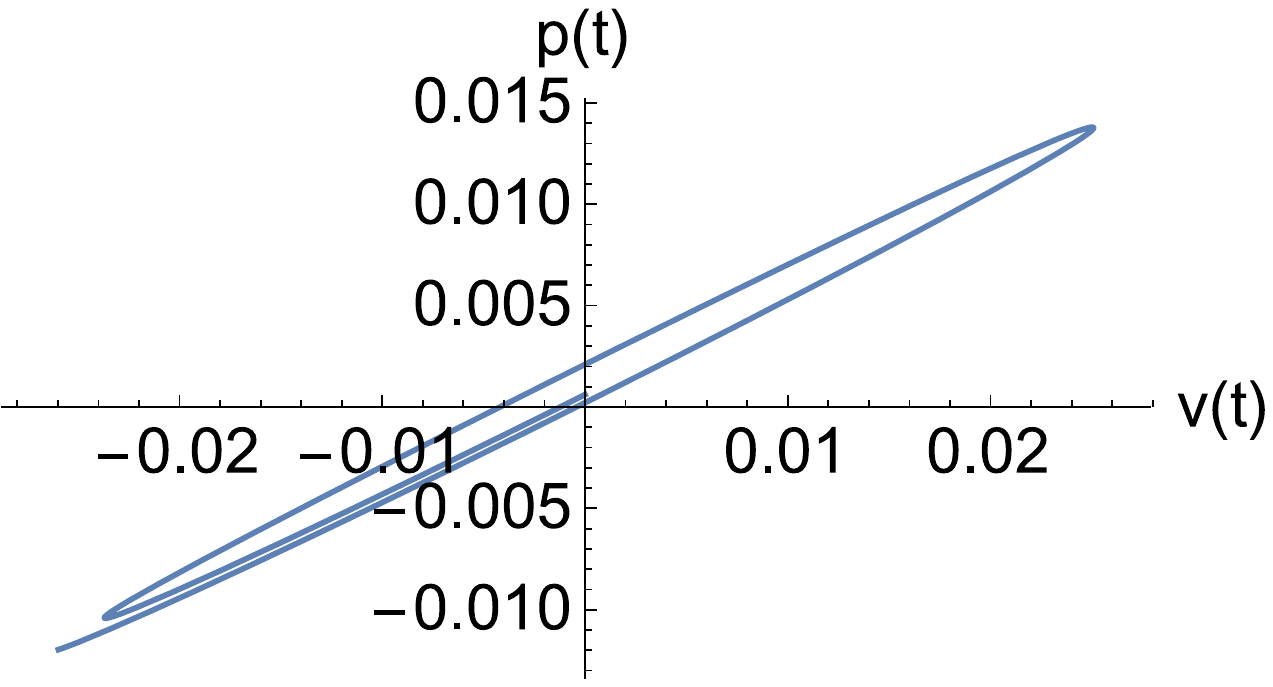}
 \caption{Trapped solitary wave when $\Delta = -\pi/3, q_0 = 1.3 \pi$. Parametric plot of $p(t)$ vs. $v (t)$ for $0<t < 120$.  Blue curve is our 4  CC  approximation.}   \label{trapdelpv}\end{figure}

\begin{figure}[h]
 \centerline{\includegraphics[scale=0.4]{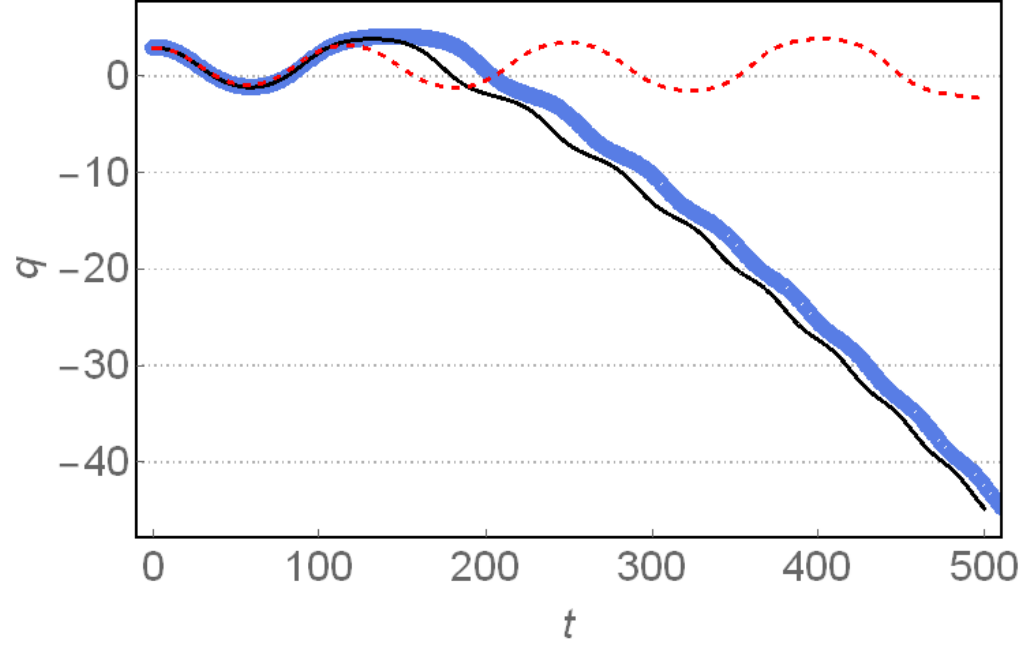}\includegraphics[scale=0.4]{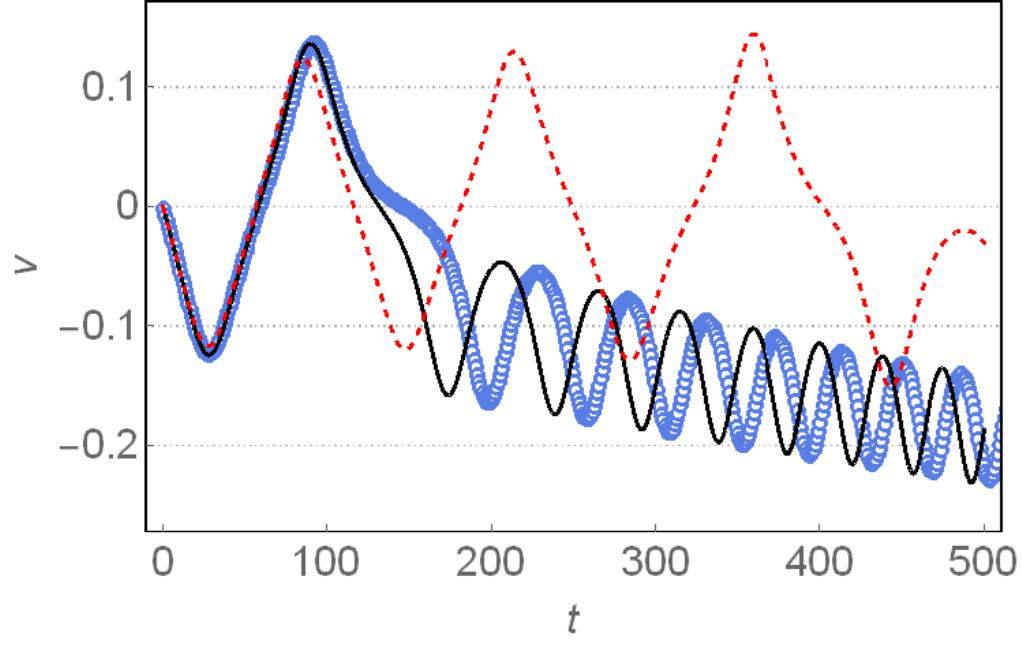}} 
 \centerline{\hspace{1cm}(a)\hspace{7cm}(b)}
 \centerline{\includegraphics[scale=0.4]{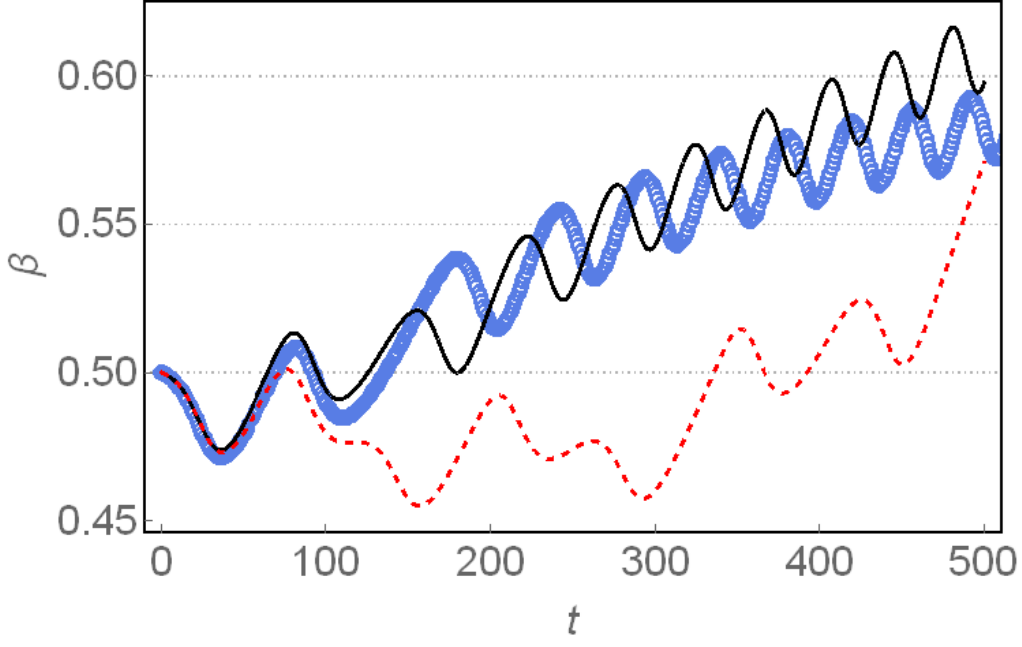}\includegraphics[scale=0.4]{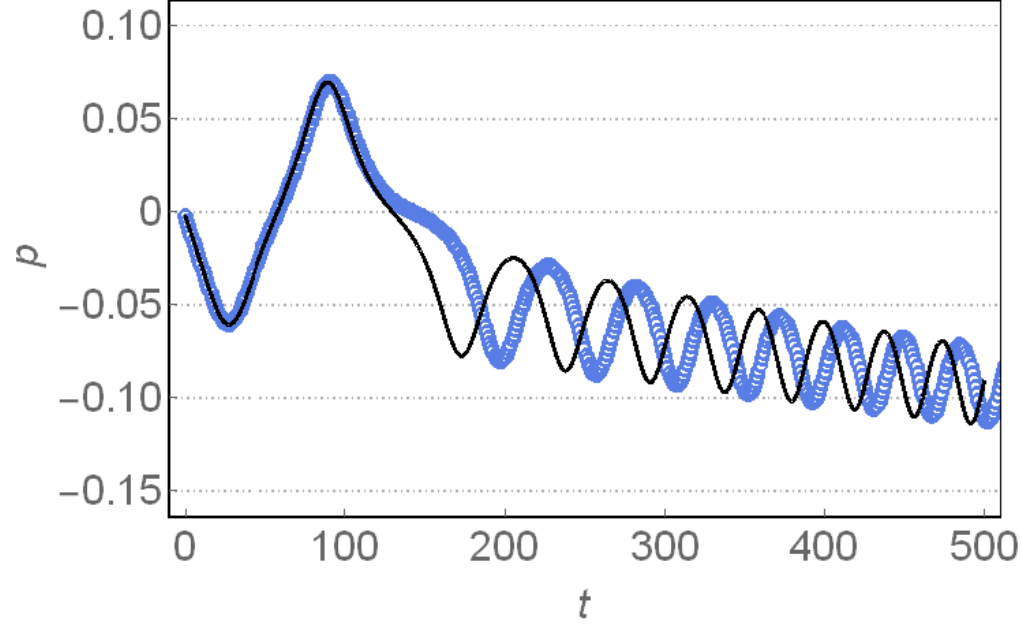}} 
 \centerline{\hspace{1cm}(c)\hspace{7cm}(d)}
 \centerline{\includegraphics[scale=0.4]{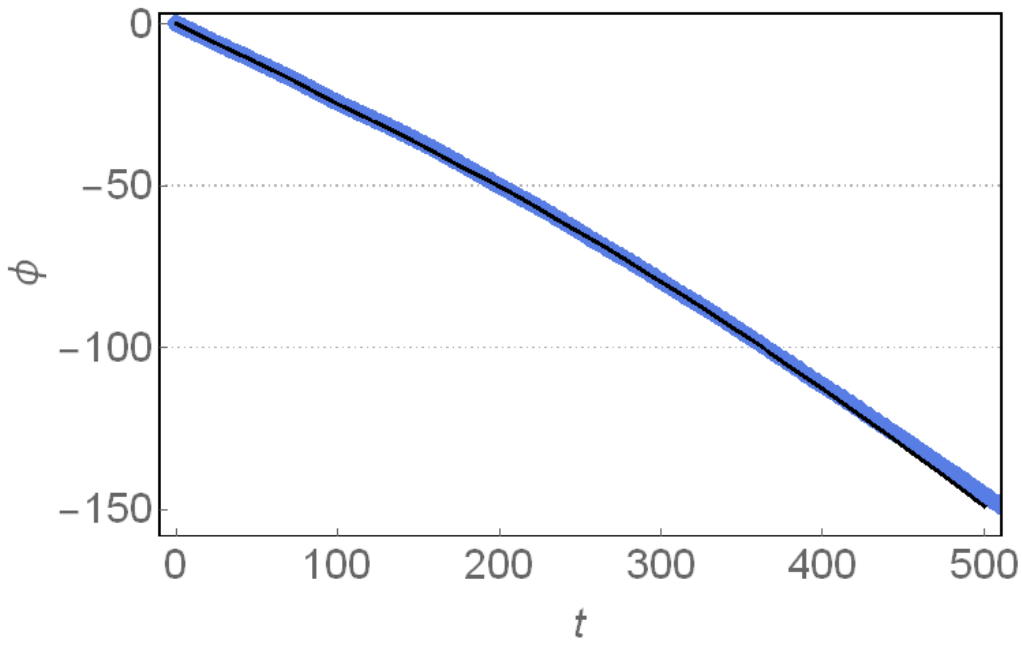}} 
 \centerline{\hspace{1cm}(e)}
  \caption{ Moving solitary wave in a shifted potential in a blowup situation: (a) position $q(t)$, (b) velocity $v(t)$, (c) amplitude $\beta(t)$, (d) momentum $p(t)$, and (e) phase $\phi(t)$.
 Here $l_0=1$,   $k_0=1$,   $V_0=-0.01$,   $W_0=-0.005$,  $\beta_0=0.5$,  $q_0=2.8\pi/3$, $v_0=0$, $p_0=0.000608946$, $\phi_0=0$, $\Delta= -\pi/3$.}
    \label{moving1}
 \end{figure}

In the next case we look at initial conditions which lead to a moving soliton whose amplitude gradually increases in time. Interestingly, the frequency of the oscillations of $v(t)$ and $\beta(t)$ only increase gradually: this is seen in Fig. \ref{moving1}. The parameters and initial conditions we choose are

\bq
V_0 = -0.01, W_0 = V_0/2, K_0 = 1, L_0 = 1;, \Delta = -\pi /3, 
q_0 = 2.8  \pi /3, v_0 = 0, \beta_0 = 0.5, \phi_0 = 0, p_0=0.000608946. 
\eq
Here we find that the 2 CC theory breaks down at around $t=100,$ whereas the 4 CC theory is qualitatively accurate for the entire time of simulation. This is seen in Fig. \ref{moving1}.

\begin{figure}[ht!]  
\centering
 \includegraphics[width=7cm]{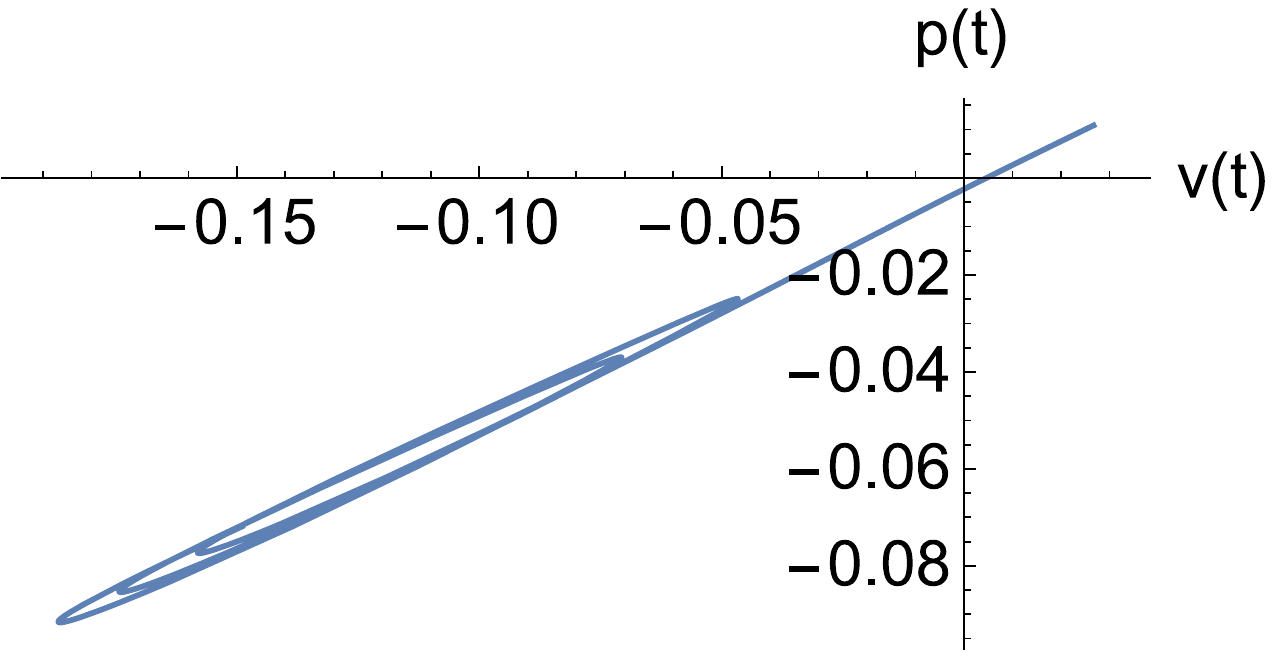}
 \caption{Moving solitary wave when $\Delta = -\pi /3, \, q_0 = 2.8  \pi /3$. Parametric plot of the momentum $p(t)$ vs. $v (t)$.  } 
 \label{movingpv}
  \end{figure}

\section{Conclusions}
We have studied the behavior of  exact solitary wave solutions of  the unforced NLSE in the presence of  complex external potentials in a collective coordinate approximation which parametrizes the wave function with four time dependent parameters.  This approximation gave excellent agreement with numerical simulations of the NLSE in most situations except the late time ``blowup" situations.  We demonstrated that our criterion for instabilities to occur, namely
$dp/dv < 0$ was a good indicator for that to happen both in our variational approximation as well as for the full numerical simulation.  We also showed that our approach was a great improvement over that of Kominis in many regimes of parameter space for various external complex potentials.  We have also demonstrated that the use of the Dissipation Functional formalism combined with a judicious choice of parametrization of the solitary wave leads to a very simple way of understanding the response of solitary waves to external complex potentials. 

\section{Acknowledgments} 
This work was supported in part by the U.S.
Department of Energy. F.G.M. \ is grateful for the hospitality of the Mathematical
Institute of the University of Seville (IMUS) and of the
Theoretical Division and Center for Nonlinear Studies at
Los Alamos National Laboratory.
 A.K. wishes to thank the  Indian National Science Academy
(INSA) for the award of an INSA Senior Professor position at Pune
University. E.A. gratefully acknowledges support from the Fondo Nacional
de Desarrollo Cient\'{\i}fico y tecnol\'ogico (FONDECYT) project No. 1141223
and from the Programa Iniciativa Cient\'{\i}fica Milenio (ICM)  Grant No. 130001.

%PGK gratefully acknowledges support from NSF-DMS-1312856,
%the Binational Science Foundation under
%grant 2010239, and the ERC under FP7, Marie Curie Actions, People,
%International Research Staff Exchange Scheme (IRSES-605096).

\end{document}